\documentclass[fleqn,10pt]{wlscirep}
\usepackage[utf8]{inputenc}
\usepackage[T1]{fontenc}
\usepackage{graphicx}
\usepackage{indentfirst}
\usepackage{subfigure}
\usepackage{dcolumn}
\usepackage{bm}
\usepackage[T1]{fontenc}
\usepackage{mathptmx}
\usepackage{etoolbox}
\usepackage{tabularx}

\title{Revealing spatio-temporal interaction patterns behind complex cities}

\author[1]{Chenxin Liu}
\author[1]{Yu Yang}
\author[1,2*]{Bingsheng Chen}
\author[1]{Tianyu Cui}
\author[1]{Fan Shang}
\author[3]{Jingfang Fan}
\author[1*]{Ruiqi Li}
\affil[1]{UrbanNet Lab, College of Information Science and Technology, Beijing University of Chemical Technology, Beijing 100029, China}
\affil[2]{Centre for Complexity Science, Imperial College London, London SW7 2AZ, UK}
\affil[3]{School of Systems Science, Beijing Normal University, Beijing 100875, China}
\affil[*]{corresponding authors: lir@buct.edu.cn (Ruiqi Li), bingsheng.chen14@imperial.ac.uk (Bingsheng Chen), jingfang@bnu.edu.cn (Jingfang Fan).}


\begin{abstract}
Cities are typical dynamic complex systems that connect people and facilitate interactions. Revealing general collective patterns behind spatio-temporal interactions between residents is crucial for various urban studies, of which we are still lacking a comprehensive understanding. Massive cellphone data enable us to construct interaction networks based on spatio-temporal co-occurrence of individuals. The rank-size distributions of hourly dynamic population of locations are stable, although people are almost constantly moving in cities and hot-spots that attract people are changing over time in a day. A larger city is of a stronger heterogeneity as indicated by a larger scaling exponent. After aggregating spatio-temporal interaction networks over consecutive time windows, we reveal a switching behavior of cities between two states. During the ``active'' state, the whole city is concentrated in fewer larger communities; while in the ``inactive'' state, people are scattered in more smaller communities. Above discoveries are universal over three cities across continents. In addition, a city stays in active state for a longer time, when its population grows larger. And spatio-temporal interaction segregation can be well approximated by residential patterns only in smaller cities. In addition, We propose a temporal-population-weighted-opportunity model by integrating time-dependent departure probability to make dynamic predictions on human mobility, which can reasonably well explain the observed patterns of spatio-temporal interactions in cities.
\end{abstract}

\begin{document}

\flushbottom
\maketitle

\textbf{The proper functioning of cities depends on efficient interactions between individuals, thus revealing universal laws underlying spatio-temporal contact patterns among residents is essential for various urban studies, including epidemic intervening, urban planning, and traffic engineering. 
From high resolution spatio-temporal networks based on co-occurrence of individuals, we discover that the Zipfian distributions of dynamic population of each location in each hour are very stable and following similar patterns over cities across continents, which suggests that cities are of similar general agglomeration patterns over time in a day. 
After aggregating spatio-temporal interaction networks in consecutive unit time windows, through the lens of community detection, we find that urban systems switch between ``active'' and ``inactive'' states, which are of contrasting interaction patterns. And larger cities have a shorter inactive time, which also indicate a faster pace of life in larger cities that is consistent with urban scaling theory. 
Moreover, spatio-temporal interaction patterns can be applied to make more comprehensive evaluation on segregation, and we find that residential patterns in small cities can better reflect interaction segregation but not in larger cities. 
All these findings contribute to gaining deeper understandings of emergent collective interaction patterns of cities and aiding better managements of urban systems.}

\section*{Introduction}

The city is a typical complex system that connects people and facilitates interactions between its residents \cite{sim2015great,west2018scale}.  
Nonlinear and diversified interactions induced by agglomeration effect with supports from more efficiently utilized infrastructures in cities \cite{li2017simple,bettencourt2013origins} give rise to the increasing return to scale, which can be depicted as a super-linear urban scaling relation between socioeconomic output and urban population \cite{bettencourt2007growth}. 
Diverse interactions from different social groups are critical for improving social cohesion \cite{alfeo2019assessing}, overall safety level \cite{logan1987racial}, well-beings of residents \cite{acevedo2003residential,toth2021inequality}, and the prosperity of cities
\cite{jacobs2016death}. 
Urban geography, social networks, and mobility of residents are crucial for increasing the diversity of interactions \cite{toth2021inequality}. 
In addition, human interaction activities also exhibit regular patterns \cite{song2010modelling,schlapfer2021universal,alessandretti2020scales} and strong rhythms that one generally wake up, go to work, back home to rest and eat, or go for entertainments \cite{west2018scale}. 
A city is composed of people and shaped by collective human interactions, 
however, we are still lacking a clear picture of interaction patterns at urban scale with a high spatio-temporal resolution. 
New York, Hong Kong, and Shanghai are among the first a few ``sleepless'' cities in the world, where interactions are seemingly happening everywhere and every moment. 
Behind almost non-stop traffic flows, are there any stable and general patterns of human concentration and interactions in cities? Is a city really sleepless? Is the city center an exceptional aggregator of interactions and diversity according to classical urban theories \cite{pan2013urban,smith1776wealth}? 
Gaining a deeper understanding of spatio-temporal patterns of human interactions in cities is crucial for answering above questions and benefiting various urban studies, including intervening epidemic spreading \cite{balcan2009multiscale,schlapfer2014scaling,li2017effects,deville2016scaling,li2020early,oliver2020mobile,li2018effect,zhong2021country,mo2021modeling}, assisting better urban planning \cite{batty2013new}, evaluating segregation \cite{louf2016patterns,chodrow2017structure,wang2016daily,xu2019quantifying}, traffic engineering \cite{helbing2001traffic,de2011modelling,li2021gravity,qiu2022understanding,li2022emergence}, resources allocation \cite{ruan2020dynamic}, and emergency management \cite{lu2012predictability,bagrow2011collective}.  

However, measuring spatio-temporal human interactions at a fine resolution was technically challenging and costly or even impossible at a large scale just a few decades ago \cite{eagle2006reality} until the emergence of massive cellphone data \cite{blondel2015survey}. 
Mobile phones can be regarded as ubiquitous sensors of various human activities since they are of a very high penetration rate and a lower usage bias across the world \cite{blondel2015survey}. 
Call and data detailed record (CDDR) of cellphone contains information about both timestamps and locations when anonymized users use the service (e.g., making phone calls, sending texts, or using data by APPs), which is generally of high spatio-temporal resolutions \cite{blondel2015survey}. For example, in some megacities, the spatial resolution of CDDR can be around two hundred meters, and down to tens of meters in the downtown area even around ten years ago \cite{xu2017clearer,xu2019unravel}. After fully upgrading to 4G or 5G networks, the spatio-temporal resolutions of CDDR can be much higher. 
Therefore, CDDR can be used to infer potential spatio-temporal interactions and mobility patterns of individuals at a fine scale.

In this study, by exploiting massive cellphone data of three cities across continents -- Dakar \cite{dong2016population}, Abidjan \cite{li2017effects}, and Beijing \cite{xu2017clearer,xu2019unravel}, we construct interaction networks based on the spatio-temporal co-occurrence of individuals \cite{mo2021modeling} to study the emergent collective patterns at the urban scale. 
Though people are almost constantly moving in cities and hot-spots that attract people are changing over time in a day, at each snapshot, Zipfian rank-size distributions of hourly dynamic population of locations is stable.  
After aggregating spatio-temporal interaction networks of two or more consecutive time windows, we observe that cities are switching between ``active'' and ``inactive'' states. 
During the ``active'' state, the population is largely concentrated in 
fewer larger communities; while in the ``inactive'' state, the population is scattered in 
more smaller communities. The above discoveries are universal over three cities across continents. 
When a city grows larger in terms of population, its ``inactive'' state shortens. 
In addition, we discover that spatio-temporal interaction segregation can be well approximated by residential patterns only in smaller cities (see Note 1 in the supplementary material).  
We further propose a temporal population-weighted opportunity (TPWO) model by incorporating time-dependent departure probability, and it can well explain the observed patterns of spatio-temporal interactions in cities.

\section*{Results}
\begin{table*}[htp]
\resizebox{\columnwidth}{!}{\begin{tabular}{ccccccccccc}

 \hline\hline
City    & Pop. &  Area & Temp. & RH & Daylight & GDP p.c. &\#Towers & \#Users & Resolution & Month          \\ 
\hline
Dakar (Senegal)   & 2.62 M      & 202  $\mathrm{km}^2$ & 22$^\circ C$ & 58\% & 11.3h      & \$ 679 & 453                & 59,864                                   & 611 $\mathrm{m}$  & Jan., 2013    
\\
Abidjan (C\^ote d'Ivoire) & 4.06 M                   & 588 $\mathrm{km}^2$  & 28$^\circ C$ & 84\%  & 11.8h      & \$ 1,132  & 242                & 12,880                                   & 1,185 $\mathrm{m}$  &  Dec., 2011      
\\
Beijing (China) & 17.7 M                 & 16,330 $\mathrm{km}^2$ & -9$^\circ C$ & 64\% & 9.5h       & \$ 14,745  & 17,317              & 90,500                                   & 177 $\mathrm{m}$   & Dec., 2013             \\ 
  \hline\hline
\end{tabular}}
\caption{\label{tab:citiesStats}Basic statistics of cities and cellphone data. RH refers to relative humidity. Area refers to the studied area with coverage by cellphone towers, for Dakar and Abidjan, the area of their administrative region is 544 $\mathrm{km}^2$ and 1,965 $\mathrm{km}^2$, respectively. Basic statistics of cities are subject to the time period of the dataset. The temperature, humidity, and length of daylight are the average of the month. \#Towers refers to the number of cellphone towers in the city related to the dataset. \#Users refers to the number of active users used in this study after filtering. 
Resolution refers to the median diameter of Voronoi polygons that are generated based on the spatial distribution of cellphone towers (see Fig. S1 and Note 1 in the supplementary material). 
}
\end{table*}

\subsection*{\label{sec:level2}Data}

To construct individual spatio-temporal interaction networks in cities and reveal possible universal patterns behind them, we exploit massive cellphone data of three cities across continents -- Dakar \cite{dong2016population}, Abidjan \cite{li2017effects}, and Beijing \cite{xu2017clearer,xu2019unravel}. 
These cities are of different urban size, climate, economic development level (see Table \ref{tab:citiesStats}), and geography (see Fig. S1 in the supplementary material). Beijing, the capital of China, is an international mega city that accommodates more than twenty million people nowadays. Abidjan, the economic center and former capital of C\^ote d'Ivoire, is one of the prosperous cities in Africa. It is also the financial and trade center of West African. Dakar, the capital of Senegal, is relatively underdeveloped.  

Call and data detail record (CDDR) of cellphone data contains both location coordinates and timestamps when anonymized customers use the service (e.g., making phone calls, sending texts, or using data by APPs), thus provides rich spatio-temporal information about human mobility patterns. Each CDDR record contains an anonymized user ID, start time, end time, latitude and longitude of the location. 
The Beijing dataset contains 100,000 subscribers and covers the whole month of December, 2013. After filtering less active users, we end up with 90,500 users (see Table \ref{tab:citiesStats}). The spatial resolution is quite high for the Beijing dataset indicated by a small median diameter of Voronoi polygons partitioned based on cellphone towers (see Table \ref{tab:citiesStats}). 
The Abidjan dataset is extrated from the one detailing high-resolution mobility trajectories of 50,000 randomly sampled users for two weeks in C\^ote d'Ivoire \cite{blondel2012d4d}, after filtering, we end up with 12,880 active users with their home located in Abidjan. 
The Dakar dataset is obtained in the same way from dataset for the whole country (Senegal) that covers two weeks for about 300,000 randomly sampled users \cite{de2014d4d}, and we end up with 59,864 active users residing in Dakar. 

\subsection*{Constructing spatio-temporal interaction networks}

\begin{figure*}[!htbp] 
\centering
\includegraphics[width=1\textwidth]{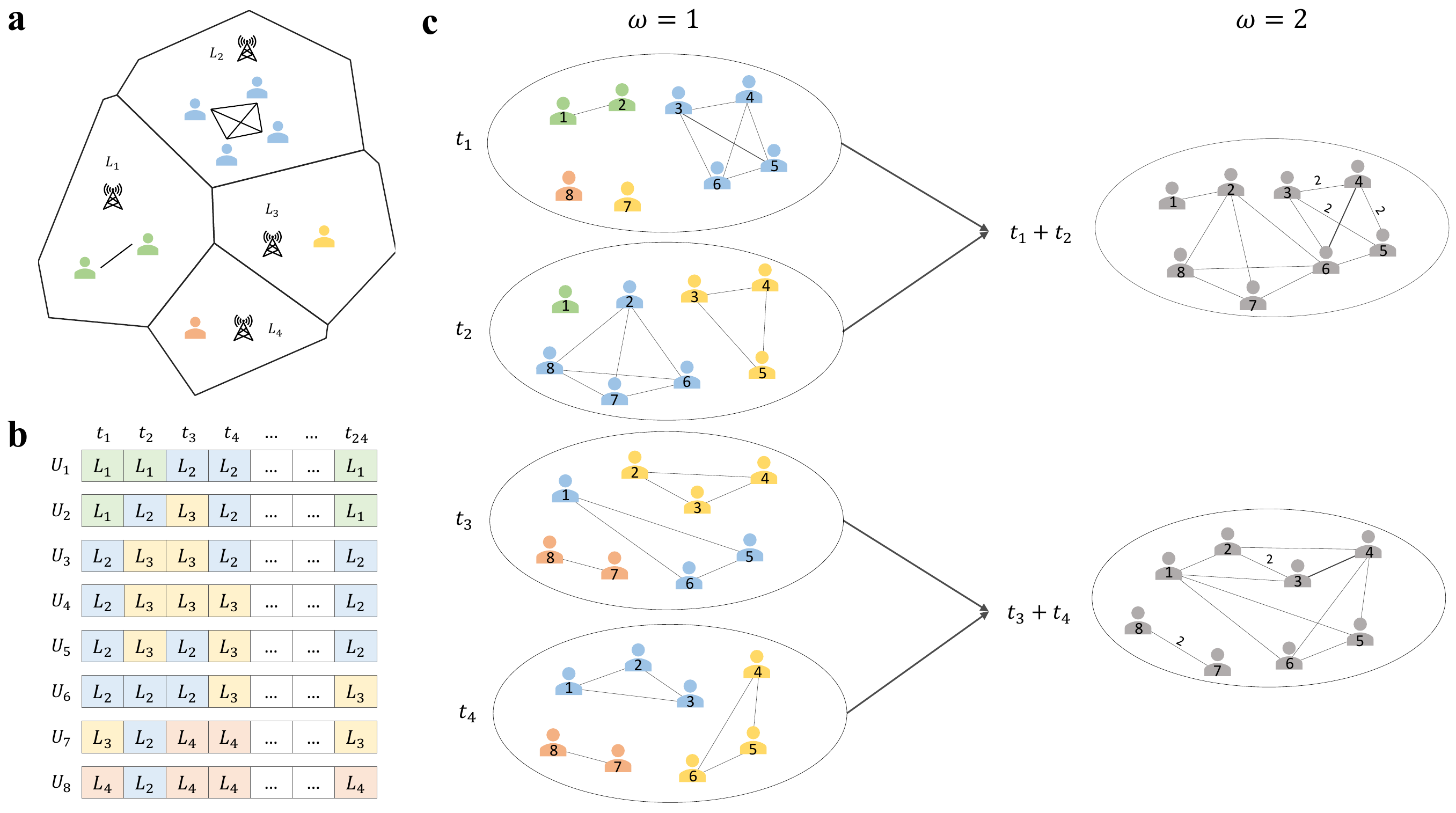}
\caption{\label{fig:space_time}Schematic diagram of constructing spatial-temporal interaction networks from cellphone data and network aggregation process. (a) A interaction network is created for each hour based on spatio-temporal co-occurrence. Individuals who appear at the same location at the same $t$ form a clique, as, for simplicity, we assume the interacting probability is equal to one.  
(b) The mobility trajectory of individuals. Locations are obtained from Voronoi tessellation based on the spatial distribution of cellphone towers (see Fig. S1 in the supplementary material). In each time window $t$, the individual is associated with the location with the longest stay duration (see Note 1 in the supplementary material).   
(c) Spatio-temporal interaction networks obtained from (b) at each snapshot (i.e., time window size $\omega=1$) and the aggregation of spatio-temporal networks over consecutive time windows with $\omega=2$. When aggregating network of snapshots (e.g., $t_1$ and $t_2$), links accumulate. 
}
\end{figure*}

Based on the CDDR data, following the framework outlined in refs.  \cite{alexander2015origin,ccolak2015analyzing,xu2017clearer,xu2019unravel}, we extract the mobility trajectory of users at high spatio-temporal resolutions (see Note 1 in the supplementary material). The obtained trajectory tells us when and where the user is, thus individual interaction networks can be inferred based on the spatio-temporal co-occurrence. If users appear at the same place and the same time, then these users will be connected and form a clique (i.e., a complete sub-graph where every two distinct nodes are adjacent) in the  spatio-temporal interaction network of that time window (see illustration in Figure  \ref{fig:space_time}). 
The real spatio-temporal interaction probability between people is affected by the land use type of the location. 
For simplicity and without losing generality, we 
make a homogeneous assumption that 
the chance of interaction between individuals at all locations are the same and equals one  
which can be suitable for COVID-19 pandemic spreading modeling \cite{STcompanions,oliver2020mobile}. And a smaller interaction probability for individuals does not change the discovered collective interaction patterns (see Fig. S11 in the supplementary material).
The highest spatial resolution we can obtain is at the location level partitioned according to the spatial distribution of cellphone towers by Voronoi tessellation (see illustration in Figure \ref{fig:space_time}a and Fig. S1 in the supplementary material). The temporal resolution depends on the frequency of cellphone usage, which can be quite high in the era of 4G and 5G communication. 
Here, we set it as one hour and associate the location with the longest stay duration to the user in each hour, and the full individual mobility trajectory consists of twenty-four locations (see Figure \ref{fig:space_time}b). Note that a smaller time window that can guarantee meaningful interactions (e.g., half hours) does not change discovered collective patterns (see Fig. S12 in the supplementary material).  


Due to mobility of individuals (see Figure \ref{fig:space_time}b), the spatio-temporal network in each time window can be varying (see the left side of Figure \ref{fig:space_time}c). 
Aggregating interaction networks over consecutive time windows connects the separated cliques in each snapshot forming more complex structures (see the right side of Figure \ref{fig:space_time}c). A prominent one is the community structure with more internal connections than external ones, which corresponds to groups of individuals who have more spatio-temporal overlaps with each other.   
Thus the community structure is a natural way to measure the collective interaction patterns at the urban scale \cite{nie2021understanding}. 
The mathematical definition of spatio-temporal networks is summarized in Appendix A. 

\subsection*{Stable rank-size distributions of hourly dynamic population}

\begin{figure*}[htb!] \centering
\includegraphics[width=1\textwidth]{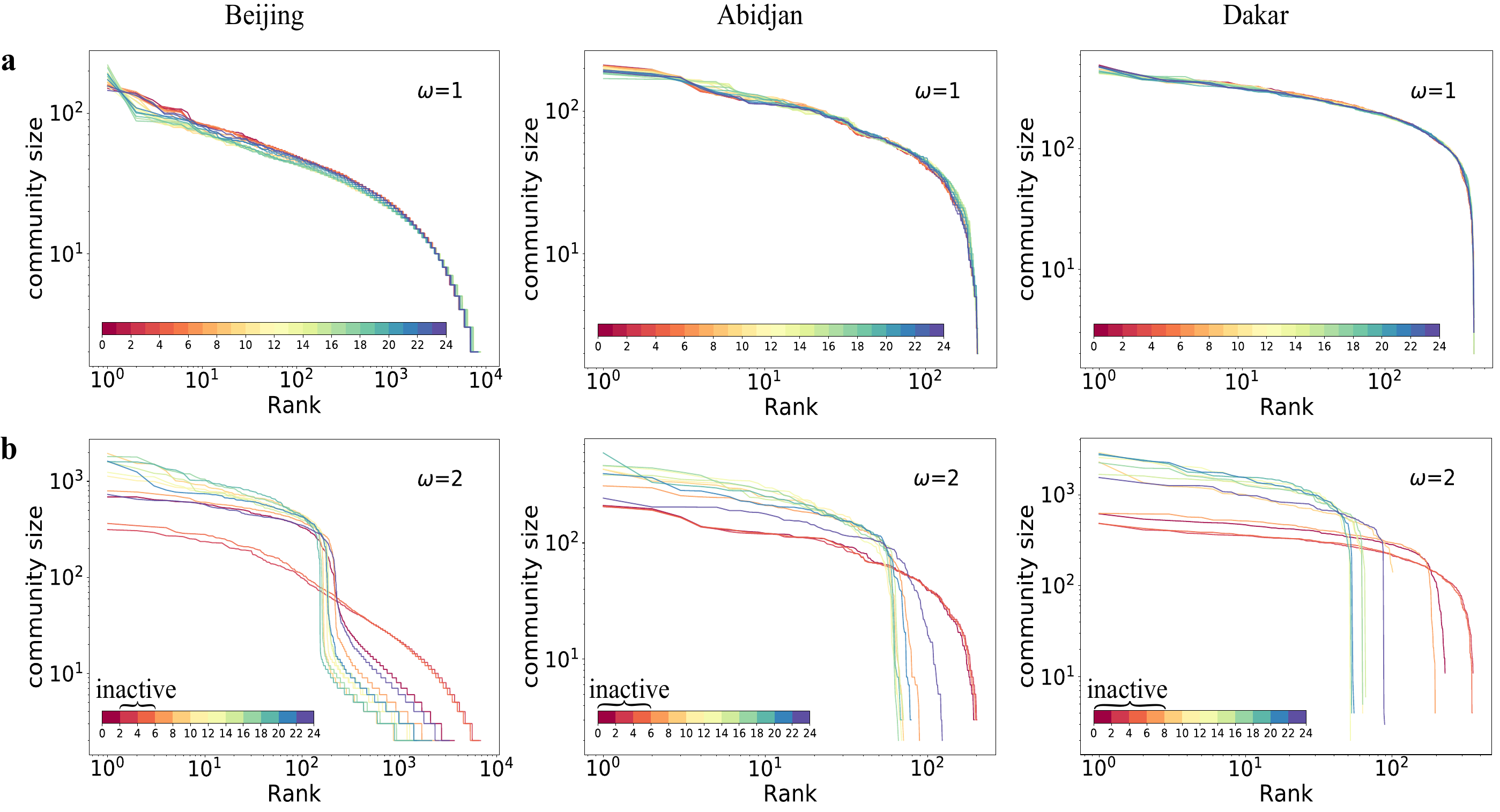}
\caption{The Zipfian rank-size distribution of communities with different time window sizes. (a) When time window size $\omega=1$, rank distributions of community size, which can be interpreted as hourly dynamic population, almost collapse together in all three cities. Given the fact that people are almost constantly moving in cities, the stability of such distributions on population concentration is non-trivial. The scaling exponent $\alpha$ of Beijing, Abidjan, and Dakar are 0.23, 0.17, and 0.09, respectively (see Fig. S2 in the supplementary material).  (b) When $\omega=2$, a switching between active and inactive states is observed across cities, which is depicted by two distinguished distributions. 
This can be easily identified via visual inspections, and more quantitative evaluation can be found in Methods and Fig. S6 in the supplementary material. 
In the active state, the city concentrates in fewer larger communities, while in the ``inactive'' state, it scatters in more smaller communities. In addition, when a city grows larger, it stays in the active state for a longer time. 
The community structure is detected by the Louvain algorithm \cite{blondel2008fast}. The color of the line corresponds to different time periods as indicated in the color bar. There are 24 lines in (a) that correspond to networks in each hour, and 12 lines in (b), each of which aggregates two consecutive networks in (a).}
\label{fig:switch}
\end{figure*}

In a specific spatio-temporal network, the network is composed of disconnected cliques (see Figure \ref{fig:space_time}a and left side of Figure \ref{fig:space_time}c), and community detection algorithm returns such divisions (see Figure \ref{fig:switch}a). The size of cliques in each location can be interpreted as the dynamic population \cite{deville2014dynamic} of that place at that time. 
To earn their livelihood and meet various needs, people are almost constantly moving in cities \cite{barbosa2018human,west2018scale}.  Hot-spots that attract most people evolve over time in a day (see Fig. S3 in the supplementary material), for instance, during lunch time, hot-spots would move to restaurants, while, during working time, the hot-spots might move to CBD and working zones. 
Thus, it is intriguing that the distribution of hourly dynamic population are fairly stable over time across cities (see Figure \ref{fig:switch}a), which might be induced by heterogeneous population distribution and temporal heterogeneity of mobility. 
Communities are ranked according to their sizes, where the largest one is ranked first and the smallest ranked the last. 
We find that such Zipfian rank-size distributions of community size of each hour almost collapse together and follow a truncated power-law 
\begin{equation}
    s \propto{r(s)^{-\alpha}e^{-r(s)/\kappa}},
\end{equation}
where $s$ is the community size, $r(s)$ is the rank of the community, $\alpha$ is the scaling exponent, and $\kappa$ is the exponential cut-off, beyond which the exponential term dominates (see Fig. S2 in the supplementary material for more details). The scaling exponent $\alpha$ is larger for cities with more population ($\alpha_{Beijing}=0.23$, $\alpha_{Abidjan}=0.17$, and $\alpha_{Dakar}=0.09$), which indicates that a larger city has a stronger heterogeneity and diversity, i.e., population concentrate in fewer larger communities and the size of communities decays faster (see Fig. S2 in the supplementary material). 





\subsection*{Switching between interaction states at urban scale} 

When aggregating spatio-temporal networks of consecutive time windows (see Figure \ref{fig:space_time}c, Figure \ref{fig:switch}b, and Fig. S2-4 in the supplementary material), the mobility of individuals are implicitly incorporated and an intriguing collective behavior emerged (see comparisons between Figure \ref{fig:switch}b and Figure \ref{fig:switch}a). Though the rank distributions of hourly dynamic population are stable, cities manifest a switching behavior between 
``active'' and ``inactive'' states, which are depicted by two distinguished rank distributions of community size (see Figure \ref{fig:switch}b and Methods for more details). In the active state, people in the whole city are concentrated in fewer larger communities, which indicate that stronger interactions are existing between people. While in the ``inactive'' state, the city is scattered in more smaller communities (see Figure \ref{fig:switch}b). 
Such discoveries are general across cities and are stable over different time window sizes $\omega$ (see Fig. S3-6 and Note 2 in the supplementary material).  
In addition, when a city grows larger (e.g., from Dakar to Abidjan to Beijing), the city stays in the active state for a longer time. This is consistent with the observation that the pace of life is faster in larger cities \cite{bettencourt2007growth}. 
For instance, Dakar stays ``inactive'' from 0 am to 8 am, Abidjan from 0 am to 6 am, and Beijing from 2 am to 6 am in their local time. 
We find that the population is the most deciding factor compared to other geographical, climatic factors, daylight length, temperature, and so on (see Table \ref{tab:citiesStats}). Although the average temperature of Beijing is much lower than the other two, it is still the most active city among them.   
Noting that all results presented in Figure \ref{fig:switch} are obtained from data within one day, in other days, all discoveries hold well (see Fig. S8-10 and Note 3 in the supplementary material). 
In addition, such discoveries on collective behaviors patterns are robust when the spatio-temporal interaction probability of individuals is not one (see Fig. S11 in the supplementary material) and when the unit window size is smaller (see Fig. S12 in the supplementary material). 

\subsection*{The relation between spatio-temporal interaction segregation and residential patterns} 


\begin{figure*}[htb!] \centering
\includegraphics[width=1\textwidth]{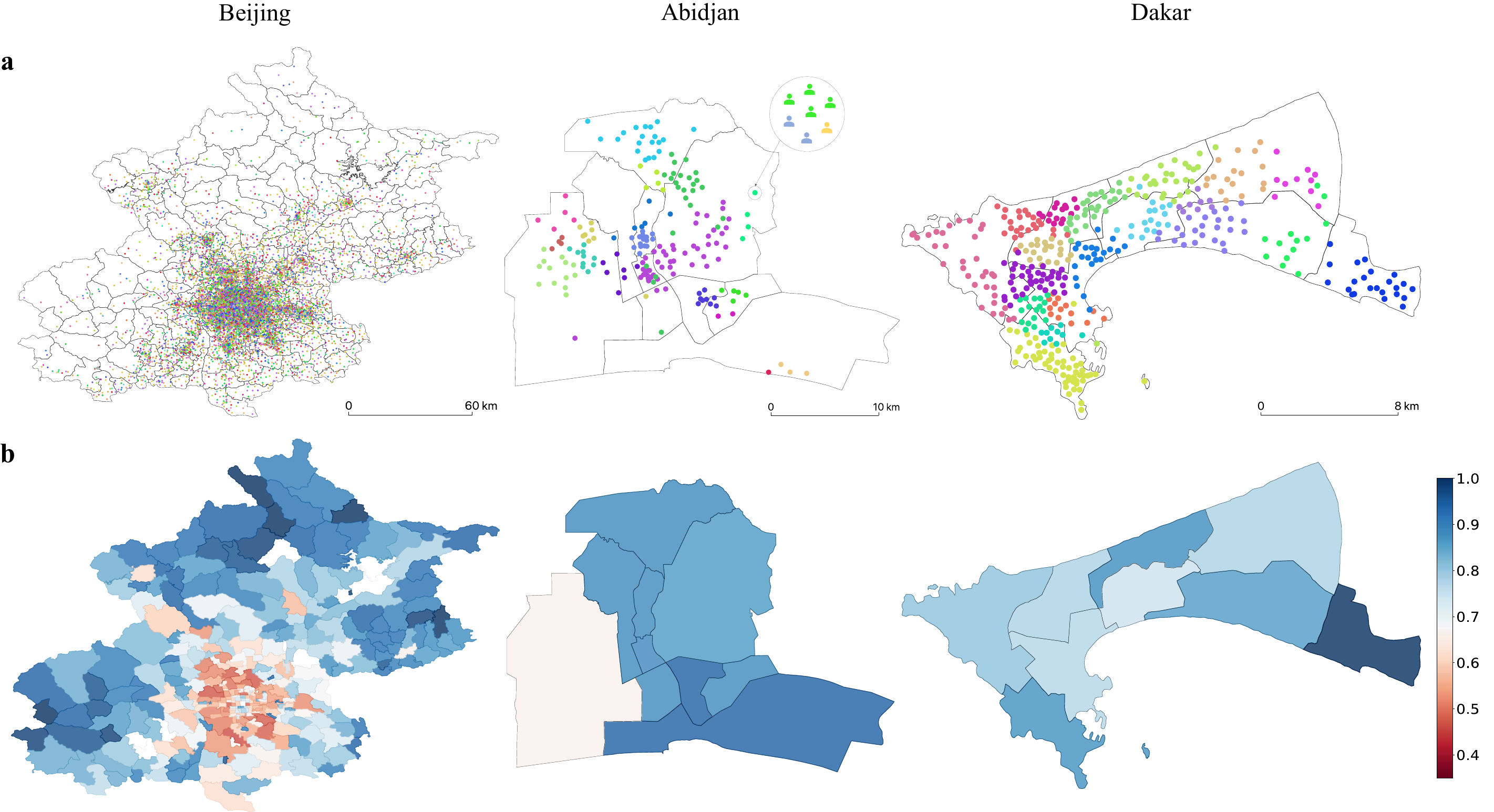}
\caption{
Empirical results on the relation between spatio-temporal interaction segregation and residential patterns. (a) The spatial distribution of interaction communities based on home locations of community members. 
The label of a location is determined by the majority rule (see the Inset in the middle). Different community labels are denoted by different colors.  Each dot denotes a location. 
In smaller cities (see Abidjan and Dakar), residential patterns can well approximate the spatio-temporal interaction segregation, but not in the larger city (see Beijing). (b) The segregation index of each \textit{Jiedao} in Beijing, \textit{Commune} in Abidjan, and \textit{Arrondissements} in Dakar. The average diameter of these administrative boundaries is around 7.96 km, 8.65 km, and 5.25 km, respectively. 
} 
\label{fig:segregate}
\end{figure*}

In previous literature, due to limitation on data accessibility, it is impossible to comprehensively study interaction segregation, and most work regarded the residential segregation as a proxy \cite{louf2016patterns,chodrow2017structure}. 
Residential segregation assumes that the home location of individuals impacts 
interactions with different groups of people (e.g., ethnicity, income) \cite{moro2021mobility,wang2016daily,gao2022quantifying}. Yet the relation between residential and interaction segregation has not been well studied \cite{xu2019quantifying}. 
Compared to social experiments or surveys in traditional studies, with massive cellphone data, we are able to evaluate whether the residential segregation, which can be measured from residential patterns, is representative of interaction segregation or not. 

Here, we aggregate all spatio-temporal interaction networks in a day (equivalent to set $\omega=24$, see Fig. S4-6 in the supplementary material), and the community divisions can be a good proxy for interaction segregation. 
As individuals in the same community have more overlaps in spatio-temporal trajectories and have a higher chance to interact with each other, however, interactions between individuals from different communities would be much less. 
Meanwhile, though the home location of individuals is not explicitly given by the cellphone data, it can be estimated by assuming that the residence is the location where the user has the highest appearance frequency between 10 pm and next 8 am \cite{alexander2015origin,ccolak2015analyzing,jiang2016timegeo,xu2017clearer,xu2019unravel}. 
By assigning the majority community label of residents to the location (see inset in the middle of Figure \ref{fig:segregate}a), we can then 
visualize the spatial distribution of home location of individuals in different interaction communities. 

In smaller cities (Dakar and Abidjan), communities generally locate within administrative boundaries. This indicates that individuals within the same spatio-temporal interaction community live spatially close, 
and thus residential patterns can be a good proxy of interaction segregation in smaller cities. 
However, in Beijing, there is no obvious correlation between them,  
as people who live in the same place can belong to many different interaction communities (see Figure \ref{fig:segregate}a). 
This indicates that interactions in Beijing are more spatially mixed, which might be induced by various factors, including income level, transportation efficiency, and mobility \cite{moro2021mobility,gao2022quantifying}.  
In the larger city (Beijing), 
residential proximity is not a good proxy of interaction patterns at urban scale.

Based on the fraction of different communities, we can further quantify the segregation level at a larger spatial scale (e.g., \textit{Jiedao} in Beijing, \textit{Arrondissements} in Dakar, and \textit{Commune} in Abidjan; they are also of similar spatial scale, see Figure \ref{fig:segregate}). 
By generalizing the approach in Ref\cite{moro2021mobility}, we can quantify the segregation index $s_i$ of region $i$ as 
\begin{equation}
s_i= \frac{|C|}{2|C|-2} \sum_{c \in C}\left|f_i(c)-\frac{1}{|C|}\right|,
\end{equation}
where $C$ is the set of unique community labels in the whole city, and $|C|$ equals to the number of unique community labels, $f_i(c)$ is the fraction of community label $c$ in region $i$. The segregation index $s_i$ is ranged from $0$ to $1$. When $s_i=0$, all unique communities are equally presented in $i$; $s_i=1$ corresponds to the case that only one type of community exists in the region.  
In Beijing, regions near the city center are of a higher diversity as indicated by a low segregation index, and a spatial gradient can be observed (see Beijing in Figure \ref{fig:TPWOresults}d). 
Such a pattern is less clear in Abidjan and Dakar (see Figure \ref{fig:segregate}b) due to possible impacts from various factors, including urban geography, socioeconomic development level, and population heterogeneity. 

\subsection*{Temporal population-weighted-opportunity (TPWO) mobility model}

\begin{figure}[htb!] \centering
\includegraphics[width=0.8\textwidth]{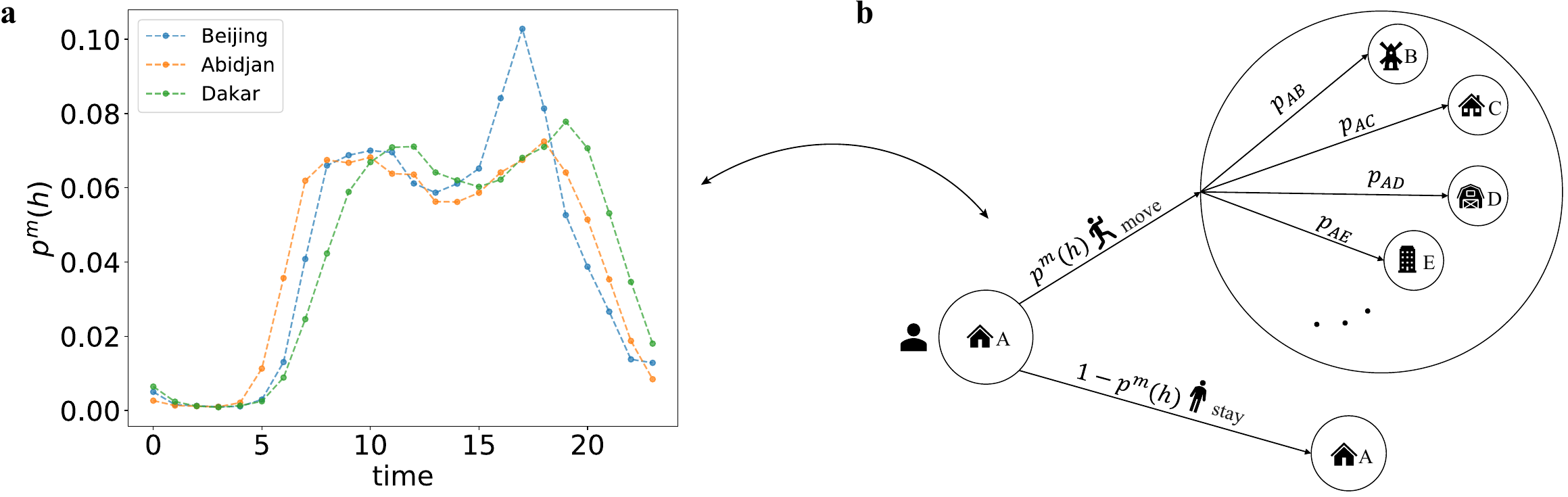}
\caption{The temporal population-weighted-opportunity (TPWO) model. (a) The departure probability of individuals estimated from massive cellphone data in cities. (b) The schema of TPWO model that a user first decide whether to move based on the departure probability $p^m(h)$ at time $h$, and if so, the probability of going to other locations (e.g., $p_{AB}$) is estimated according to Eq. (\ref{eq:PWO}).} 
\label{fig:TPWO}
\end{figure}




To better understand possible factors influencing spatio-temporal interaction patterns in cities, we model mobility trajectory of individuals based on the well-known population-weighted opportunity (PWO) model \cite{yan2014universal}. Compared to other parameter-free mobility models \cite{simini2012universal}, 
PWO is more suitable for predictions at the intra-urban scale \cite{yan2014universal}. 
However, PWO neglects the temporal evolution of mobility and is only able to make static predictions for one step. When applied to model human mobility for multiple steps, it does not work well on reproducing observed spatio-temporal interaction patterns (see Fig. S13-15 in the supplementary material). 
we propose a temporal population-weighted opportunity (TPWO) model by incorporating time-dependent departure probability $p^m(h)$, which is generally of two peaks (see Figure \ref{fig:TPWO}). And $p^m(h)$ is subject to the normalization condition that $\sum_{h=1}^{24} p^m(h)=1$. 
At each time step, the probability of movement of an individual is firstly determined by $p^m(h)$. If so, the probability of moving from location $i$ to $j$ is estimated by  
\begin{equation}
p_{ij}=\cfrac{T_{ij}}{T_i^{out}}=\cfrac{T_{ij}}{\sum_{j\neq i}^{N} T_{ij}},
\label{eq:PWO}
\end{equation}
where $T_{ij}$ is the volume of mobility flow from $i$ to $j$, and $T_i^{out}=\sum_{j\neq i}^{N} T_{ij}$ is the total outgoing flow from $i$. 
\begin{equation}
   T_{ij} = T_{i} \cfrac {m_{j}(1/S_{ji}-1/M)}{\sum\nolimits_{k\neq i}^N     m_{k}(1/S_{ki}-1/M)}, 
\end{equation}
where 
$S_{ji}$ is the population in a circle with $j$ as the center and the distance between $i$ and $j$ as the radius, $m_j$ is the population of location $j$, $T_{i}$ is generally assumed to be proportional to $m_j$ \cite{simini2012universal,yan2014universal}, $N$ is the total number of locations in the city, and $M=\sum_j m_j$ is the total population of the city.
In such settings, each individual in our model will have one trip per day on average, while in reality, people might have two to three trips \cite{wang2015beijing} (e.g., in weekdays in Beijing in 2013, it is around 2.3 trips per day per person according to travel surveys \cite{wang2015beijing}). 
From cellphone data, the estimation on the average number of trips $\langle t \rangle$ is 2.49, which is quite close to travel surveys \cite{wang2015beijing}. For the other two cities, we did not find detailed travel surveys of the year consistent with the datasets, and the estimations from cellphone data are 2.66 and 2.57 for Abidjan and Dakar, respectively.  
To make our model be more consistent with reality, at each time step, for each individual, we attempt $\langle t \rangle$ times to determine whether the individual moves or not (i.e., if he/she does not move in the first attempt, then we will try a second time, if moving, then no further attempts, otherwise, continue trying until $\langle t \rangle$ times; when $\langle t \rangle$ is not an integer, whether there will a last attempt is determined by chance according to $\langle t \rangle - \lfloor \langle t \rangle \rfloor$. For simplicity, we set $\langle t\rangle=2.5$ for all three cities in this study, which makes the number of total trips generated by the model more compatible with reality. Without such a multiplier, all results generated still hold well qualitatively (see Fig. S16 in the supplementary material). 

\begin{figure*}[htbp!] \centering
\includegraphics[width=1\textwidth]{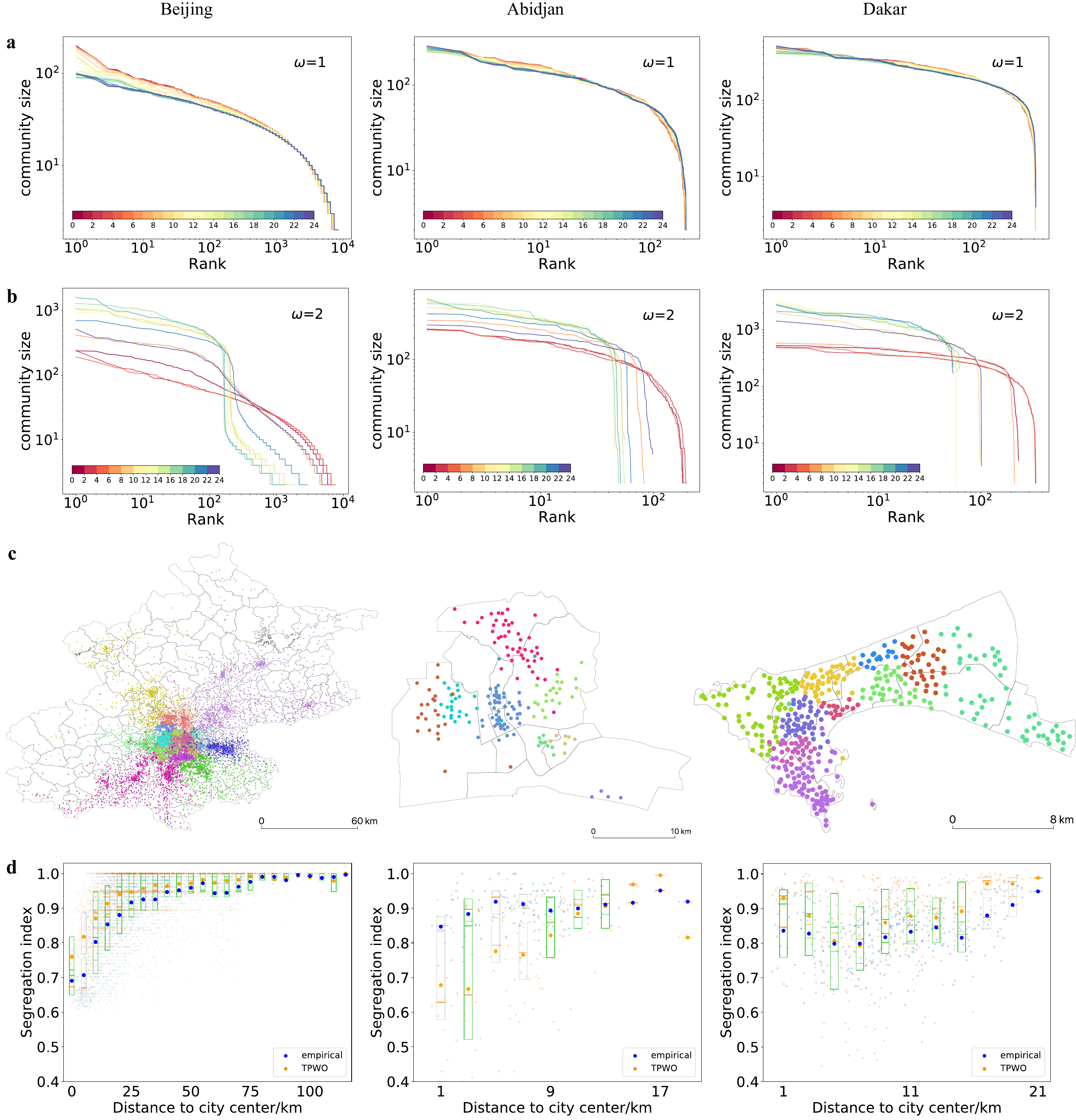}
\caption{Spatio-temporal interaction patterns generated by the TPWO model. 
Empirical results obtained from cellphone data suggest that the average number of trips of residents are 2.49, 2.66, 2.57 for Beijing, Abidjan, and Dakar, respectively. The result for Beijing is quite close to travel survey \cite{wang2015beijing}. Here for simplicity, we adjust the TPWO model with a multiplier of 2.5 for all three cities. 
Rank-size distributions of communities in spatio-temporal networks constructed from results generated by the TPWO model, when (a) $\omega=1$ and (b) $\omega=2$. (c) The spatial distribution of spatio-temporal interaction network communities generated by the TPWO model. (d) Quantitative comparisons between results generated by the TPWO model and the empirical ones from the city center to suburbs. The larger darker dots are the average, and smaller lighter dots represent results of each location. The rectangle indicates the 25\% and 75\% percentile of the data, and the darker bars in the rectangle are the median. When rectangles overlap, they are marked in green, otherwise, in grey. In Beijing and Dakar, our TPWO model can well reproduce the empirical results but is less successful in Abidjan, which might be due to relatively low spatial resolution of data and stronger spatial constraints imposed by urban terrain in Abidjan.} 
\label{fig:TPWOresults}
\end{figure*}

In our framework, we first initialize individuals to their home locations, 
and then simulate twenty-three steps of movements of each individual via the TPWO model to obtain the mobility trajectory. The size and initial spatial distribution of the simulated population are the same as users in cellphone data. From the individual mobility trajectories generated by the TPWO model, we construct the spatio-temporal interaction networks to study various interaction patterns at the urban scale. Results show that the TPWO model can well reproduce stable distributions of hourly dynamic population (see Figure \ref{fig:TPWOresults}a and Fig. S10-12 in the supplementary material) and the switching behavior between two interaction states (see Figure \ref{fig:TPWOresults}b and Fig. S10-12 in the supplementary material). 
And the residential patterns of individuals in the same interaction communities can also be well reproduced except in Abidjan (see Figure \ref{fig:TPWOresults}c,d). The possible reasons might lie in a lower spatial resolution of cellphone data (i.e., each location is of larger range than in Dakar and Beijing, see the ``Resolution'' column in Table \ref{tab:citiesStats}) and a more complex urban terrain, for example, there is a big salt lake and river wriggling in Abidjan, and a national park separates two main regions. With a lower spatial resolution, mobility within the same location is undetectable, which will make the mobility seemingly more limited in reality and will influence the resolution of calculation in Figure \ref{fig:TPWOresults}d. As in mobility model, when the spatial resolution is low, then there will be fewer locations to go in cities and a generally longer travel distance, how to overcome the problem of low spatial resolution can be closely investigated in the future.  

In addition, it is worth noting that in mobility models, population is the most critical factor that measures the attraction of a location. The commonly used proxy of attraction of locations is the residential population \cite{park2018generalized,simini2012universal,yan2014universal,yan2017universal}, which is biased to working and business zones where the residential population (RP) is usually small but those regions are of high attractions. The active population (AP) that measures the number of individuals who have visited the location in a day \cite{li2017simple} can be a much better proxy for the attraction of a location. We implement two versions of TPWO models, one is using RP, and the other is using AP of locations. TPWO model using AP has a better performance (see Fig. S13-15 and Note 4 in the supplementary material). 

\section*{Discussions}
As rapid urbanization will continue in the next a few decades \cite{bettencourt2007growth,li2017simple,feng2018evolving,li2021assessing,west2018scale}, gaining a deeper understanding of collective interaction patterns at the urban scale is crucial for inventing more livable, diversified, and prosperous future cities. In this context, our work has three useful findings. Firstly, although people are almost constantly moving in cities, the rank distribution of the hourly dynamic population is not altered by the movements of individuals among locations. Human mobility can be regarded as perturbations to the system, the discovered stable rank distributions (see Figure \ref{fig:switch}) imply that city can be regarded as a self-organized system \cite{bak2013nature,christensen2005complexity}. And bigger cities are of a higher heterogeneity on dynamic population distribution, which is urban characteristic. 
Secondly, the observed switching behavior between ``active'' and ``inactive'' states of cities is intriguing. In active states, individuals concentrate in fewer larger communities, which is similar to a reaction process that concentrates more energy and materials. While in the ``inactive'' state, the city is scattered into more smaller communities. Such discoveries can be informative on assisting urban management. For example, if a city becomes ``inactive'' at 10 pm, then, during a pandemic, setting a curfew from 11 pm will be less effective.  In addition, current findings might suggest that when a city grows larger, it stays in the ``active'' state for a longer time. When data of more cities is accessible, this can be better tested. From studies on urban scaling laws, the pace of life is faster in larger cities, and it would be interesting to investigate whether a city can be really sleepless if the population concentration continues. 
Thirdly, residential patterns can be as a good proxy of interaction segregation in smaller cities (Dakar and Abidjan). In the larger city (Beijing), there is a clear spatial gradient of segregation index from the city center to the suburbs. The city center is shared by people from different groups and is an aggregator of diversity (see Note 5 in the supplementary material). It is worth noting that we are assuming local interactions between people who live closely still exist, if this is not the case in the future, then the spatial patterns of interaction communities in Beijing  (see Figure \ref{fig:segregate}) might be the most segregated one: as people who live spatially close may not have strong interactions at all. Such a trend has already emerged in megacities, one may highly probably do not know the people who live above or below your apartment, or even just the next door. 

All these findings contribute to gaining deeper understandings of emergent collective behaviors of cities and better managements of urban systems. For example, identifying close contacts of the infected patient based on community structures in spatio-temporal interaction networks, predicting the spreading paths, and intervening the spreading of COVID-19 in cities while preserving more interactions via dismantling the spatio-temporal interaction networks. The heterogeneous distribution of population and departure probability might be two key factors influencing the collective spatio-temporal interaction patterns of cities, which is revealed by our TPWO model. 
The deeper mechanics behind them are worth closer investigations in the future.

In addition, we also bridge the collective mobility model and spatio-temporal interaction patterns of individuals. 
The PWO model is a collective mobility model that is applied to predict the volume of flows among locations but not to predict individual mobility trajectories. By incorporating heterogeneous departure probability and making a partial homogeneous assumption, which assumes that individuals at a certain place will follow the ensemble mobility probability $p_{ij}$ in Eq. (\ref{eq:PWO}) and the differences between individuals are only their home locations, the TPWO model can reasonably well reproduce collective spatio-temporal interaction patterns of cities.  
The TPWO model does not require detailed individual trajectory and other information, and only needs simple population-level inputs -- departure probability and distribution of population, which can be easily obtained. Thus our framework provides a trade-off between privacy-preserving and studying spatio-temporal individual interaction patterns at the urban scale.


\providecommand{\noopsort}[1]{}\providecommand{\singleletter}[1]{#1}%

\section*{Methods}
\subsection*{Mathematical formalization of spatio-temporal interaction networks}
Mathematically, a spatio-temporal interaction network $\mathcal{G}(t)$ is defined based on the set of spatio-temporal co-occurrence events of individuals (i.e., edges in networks) 
\begin{equation}
    \mathcal{E}(t)=\{(u,v,t)\},
\end{equation} 
where $u,v \in V$ are individuals that appear at the same location at the same time window $t$, and $V$ is the set of individuals. 
For higher accuracy, the resolution of the time window can be set quite small when the temporal resolution of the raw data is high. In this paper, for simplicity and without losing generality, we set the resolution of the time window as one hour. 
A specific spatio-temporal interaction network at time window $t$ is then formulated as  
\begin{equation}
    \mathcal{G}(t)=\{\mathcal{V}, \mathcal{E}(t)\}. 
\end{equation}
$\mathcal{G}(t)$ is an undirected network. For simplicity, we assume that $\mathcal{G}(t)$ is unweighted, where 
a non-zero $A_{uv}(t)$ represents the spatio-temporal co-occurrence of $u$ and $v$ in the time windows $t$. 
The aggregation of co-occurrence events of consecutive time windows is formulated as 
\begin{equation}
 \mathcal{E}(t,t+\Delta t)=\{(u,v,t')\},
\end{equation}
where $t'\in[t,t+\Delta t]$. 
The aggregated spatio-temporal interaction network is defined as 
\begin{equation}
 \mathcal{G}(t,t+\Delta t)=\{\mathcal{V}, \mathcal{E}(t,t+\Delta t)\}.
\end{equation}
The spatio-temporal co-occurrence events of individuals in $\mathcal{G}(t,t+\Delta t)$ accumulate. For example, if $(u,v, t=1)$ and $(u,v, t=2)$, 
then $A_{uv}(1,2)=2$ in $\mathcal{G}(1,2)$ (i.e., $\Delta t=1$, which means the time windows size $\omega=2$, see more details of the aggregation process in Fig. \ref{fig:space_time}c). 
An aggregation of networks in more consecutive time windows (i.e., $\omega >1$) generally increases overall connectivity of the graph, due to movements of individuals.


\subsection*{Quantitative detection of switching between two interaction states}
The switching between two interaction states can be easily noticed by visual inspections, and quantitatively, we find that the ``turning point'' of the curve \cite{ke2015defining} can be informative. The turning points of distributions in Fig. \ref{fig:switch}b should be the ones after which the curve decays faster. Generally, it is the farthest point from the straight line that connects the two ends of the curve (see Supplementary Fig. 6a). The rank value of turning points of distributions in working states is generally smaller than the ones of sleeping states, which gives a quantitative distinction between two states (see Supplementary Fig. 6b).

\section*{Data availability}
The cellphone data of Dakar and Abidjan is shared by organizers of Data for Development Challenges \cite{blondel2012d4d,de2014d4d}, the one of Beijing is provided by a local telecommunication operator under a non-disclosure agreement. 
The shapefile of Abidjan at the commune resolution is from ref.\cite{doumbia2018emissions}. 
Other additional data related to this paper are publicly available and may be requested from the authors.

\section*{Acknowledgements}
This work receives financial supports from the National Natural Science Foundation of China (Grant No. 61903020), Fundamental Research Funds for the Central Universities (Grant No. buctrc201825). We acknowledge organizers of Data for Development (D4D) Challenges for sharing cellphone data of Senegal and C\^ote d'Ivoire, and Dr. Madina Doumbia from University P\'el\'eforo Gon Coulibaly for providing the shapefile of Abidjan at the commune resolution. 

\section*{Author contributions statement}
R.L. conceived and supervised the research,  C.L., Y.Y., T.C., and F.S. analyzed data, C.L. constructed the mobility model, C.L., J.F., R.L., and B.C. analyzed and discussed the results. B.C., C.L., J.F. and R.L. wrote the manuscript. All authors reviewed the manuscript.

\clearpage

\section*{Supplementary Information}

\renewcommand\figurename{Supplementary Figure}
\renewcommand\tablename{Supplementary Table}

\setcounter{figure}{0}

\section*{Supplementary Note 1: Cellphone data processing}

\begin{figure}[htb!] \centering
\includegraphics[width=1\linewidth]{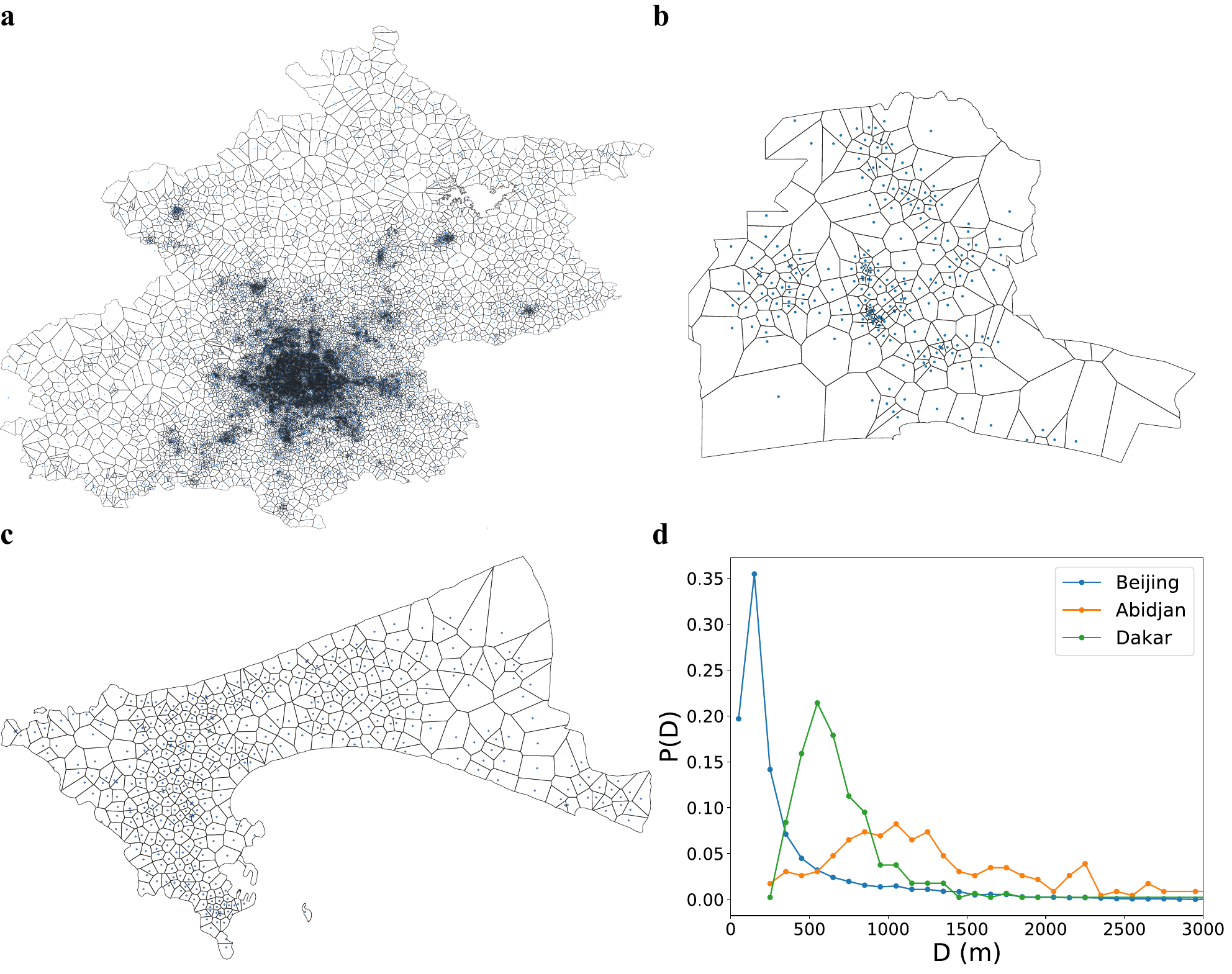}
\caption{\textbf{Spatial distributions of cellphone towers and corresponding Voronoi tessellations.} \textbf{a} Beijing, \textbf{b} Abidjan, and \textbf{c} Dakar. The spatial coverage of each tower is generated based on the Voronoi tessellation, which partitions a Euclidean plane into regions close to each cellphone tower.  
The density of towers is quite high in the central areas of cities, which is generally due to the high work load generated by the the large active population there. In general, the density of towers decays along with the increase of distance to the city center. The small blue dots in \textbf{a}-\textbf{c} represent the position of cellphone towers. \textbf{d} The distribution of the diameter of Voronoi polygons. Apart from Abidjan, more than $90\%$ of locations are of a diameter less than $1$ km.}
\label{fig:towers}
\end{figure} 

The data used in this study is the call \& data detail records (CDDR) of massive anonymized cellphone users from three diversified cities -- Beijing, Abidjan, Dakar. Cellphone data are of larger sampling rate compared to other resources, lower usage biases, low implementation cost, and quicker updates. 
Each record contains an anonymized user ID, start time, end time, latitude and longitude of the tower that the user connects to, when an individual uses some service, including making phone calls, sending texts, and using data. 

Based on the position of cellphone towers in CDDR data, we can partition the whole urban space into locations (see Supplementary Fig. \ref{fig:towers}a-c). 
The number of locations is 17,317 for Beijing, 242 for Abidjan, and 453 for Dakar. 
We find that the spatial resolution in Beijing dataset is the highest, followed by Dakar and Abidjan (see Supplementary Fig. \ref{fig:towers}d). Apart from Abidjan, more than $90\%$ of locations are of a diameter less than $1$ km. 
The size of the location is relatively small in central areas due to the high activity demands generated by the large population there. 

From CDDR data, we can estimate the mobility trajectory of individuals at the level of locations partitioned by cellphone towers. 
However, the raw CDDR data is subject to noises due to tower-to-tower call traffic balancing to improve their service by telecommunication operators. When the nearest tower is busy, the signal might be routed to a less busy tower farther away. This generates signal jumps that introduce noises, appeared as fast and long movements beyond a travel speed limit. To filter such noises, various methods have been reviewed in ref. \cite{jiang2013review}. One of the simplest yet effective methods is to remove the next record if the inferred speed between two records is beyond the reasonable speed limit. However, it heavily relies on the correctness of the first record. To improve its accuracy, we check if the first record is noisy if the speed between the first and the second record is beyond a predefined speed limit, we then remove the first record. We repeat this process until there are no artificial jumps between two records. 

The detection of the home location of users is based on a more accurate estimation of appearance frequency, when there is no ground-truth. The most visited location during weekday nights from 10 pm to next 8 am are labeled as home. This requires both a temporal clustering and a spatial clustering to obtain more accurate stay-points of users from the CDDR data after noise filtering. 
We improve upon the stay-point algorithm presented in refs. \cite{jiang2013review,zheng2011learning} as follows. (i) We apply a temporal agglomeration algorithm. The temporally consecutive records within a certain radius (e.g., 500 meters) are bundled together with an updated stay duration from the start time of the first record to the end time of the last one. Thus, when constructing spatio-temporal interaction networks, the user is associated with the location with the longest stay duration.
(ii) We then label the location as stays if the user stays there above a duration threshold (e.g., 10 minutes). We then combine all the spatially adjacent stay points for a user (within a threshold) as his or her stay regions, which will be later labeled as home, work, and other. For this spatial agglomeration, we use R-tree to accelerate the computation \cite{guttman1984r}. R-tree is a type of spatial B-tree, a spatial search balancing tree that checks the boundaries of elements to make the search faster. 
More details and the pseudo-code for the algorithm can be found in refs. \cite{xu2017clearer,xu2019unraveling}.

\section*{Supplementary Note 2: Spatio-temporal interaction networks with varying time window size $\omega$}

\begin{figure}[htb!]  \centering
\includegraphics[width=\linewidth]{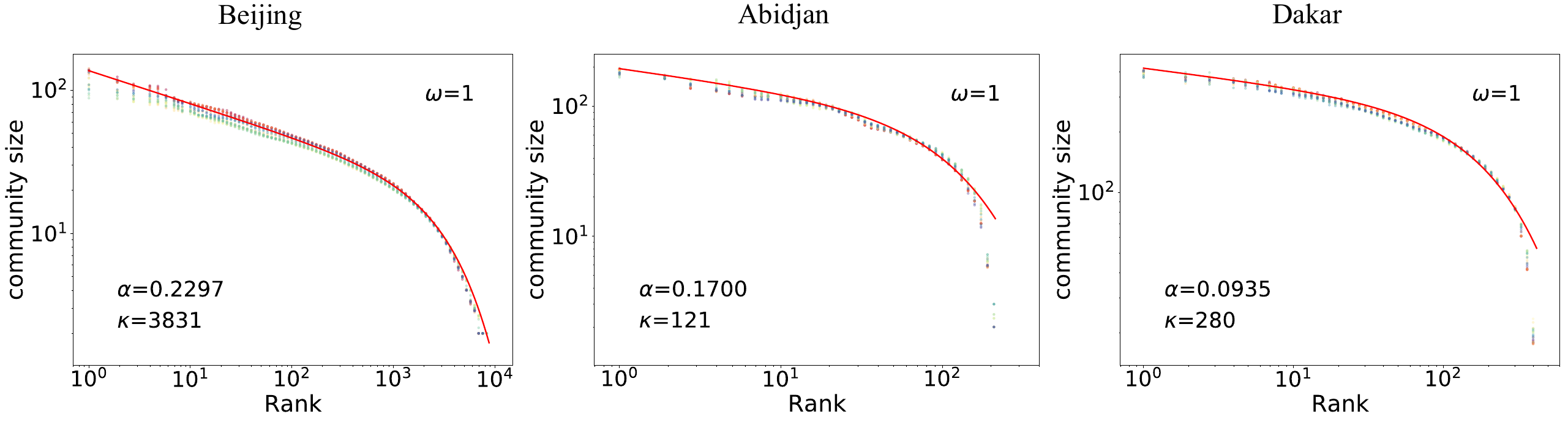}
\caption{\textbf{Zipfian rank-size distributions of community size.} When the time window size $\omega=1$, we find that the rank distribution of community size can be reasonably well fitted by a truncated power law $s \propto{r(s)^{-\alpha}e^{-r(s)/\kappa}}$, where $s$ is the size of communities, and $r(s)$ is the rank value of the community, $\alpha$ is the scaling exponent, and $\kappa$ is the exponential cut-off. $\alpha$ and $\kappa$ are labeled in the figures, and $\alpha$ is larger for cities with more population. And $\alpha$ can be regarded as an indicator on the population heterogeneity of cities.}
\label{fig:fit}
\end{figure}

\begin{figure}[htb!] \centering
\includegraphics[width=0.9\linewidth]{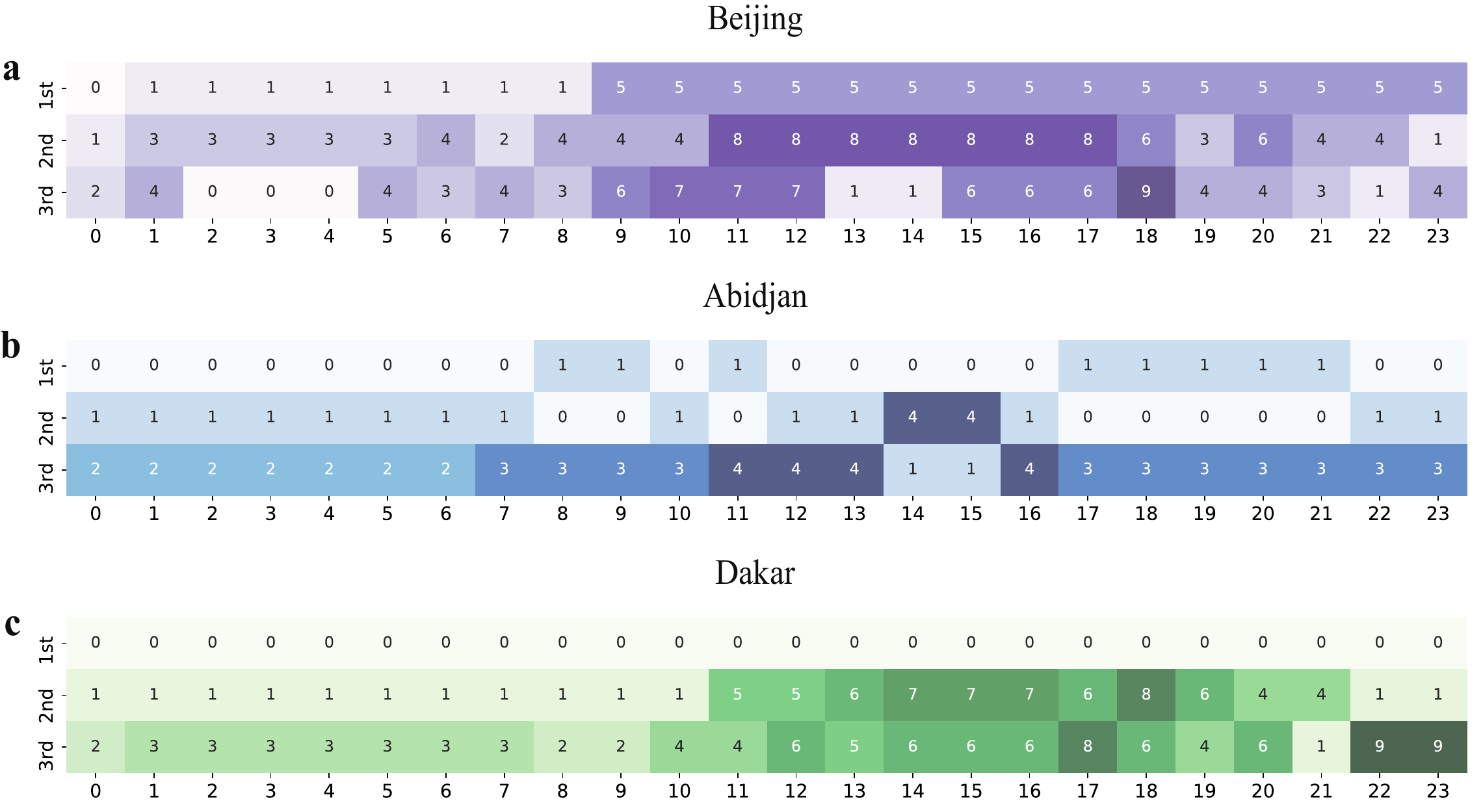}
\caption{\textbf{The top-three hot-spot locations over time in cities.} \textbf{a} Beijing, \textbf{b} Abidjan, and \textbf{c} Dakar. The x-axis denotes the hour of a day, and the number denotes a logical index of a location. The color corresponds to the value of the index, which is just an aid for better visualizing the changing of locations.}
\label{fig:hot_spots}
\end{figure}



\begin{figure}[htb!]  \centering
\includegraphics[width=1\linewidth]{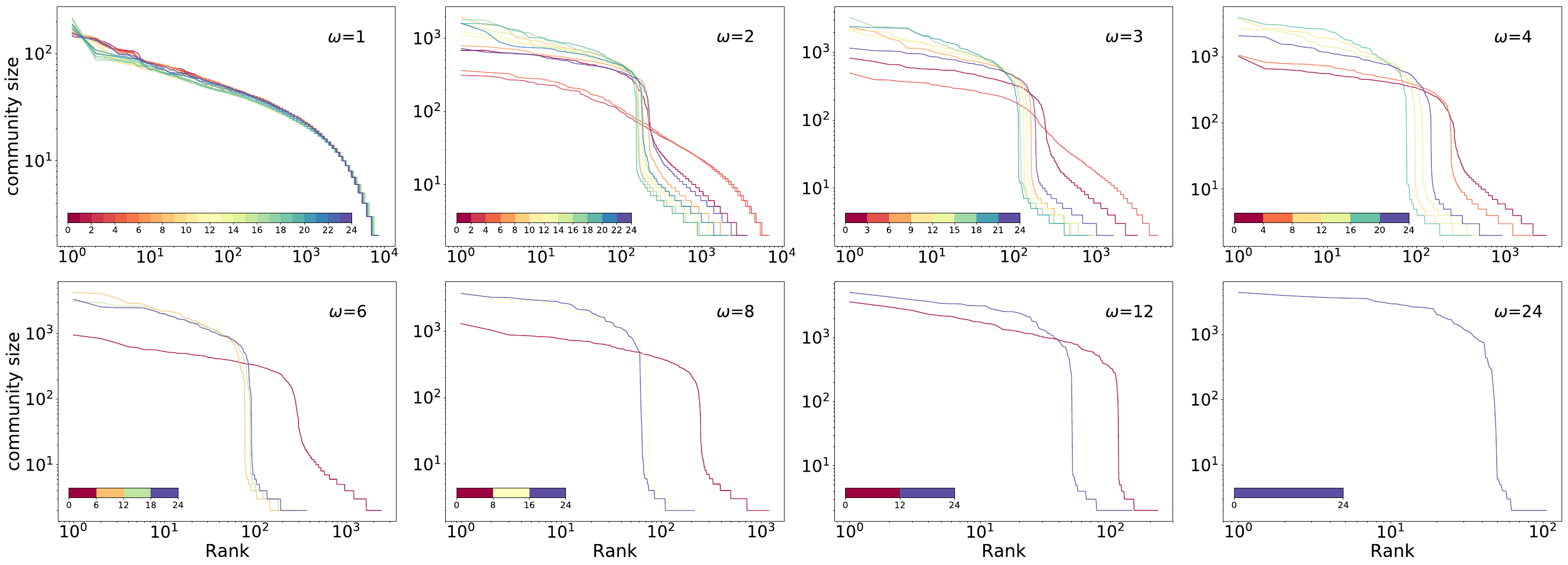}
\caption{\textbf{The Zipfian rank-size distribution of community size with varying time windows size $\omega$ in Beijing.} Each line represents the distribution of a specific network in a certain time window. When $\omega=1$, all lines almost collapse together.  
The sleeping state of Beijing is from 2 am to 6 am, which is clear from the figure with $\omega=2$, and the switching behavior always exists when $\omega\in[2,12]$.}
\label{fig:beijing}
\end{figure}

\begin{figure}[htb!] \centering
\includegraphics[width=1\linewidth]{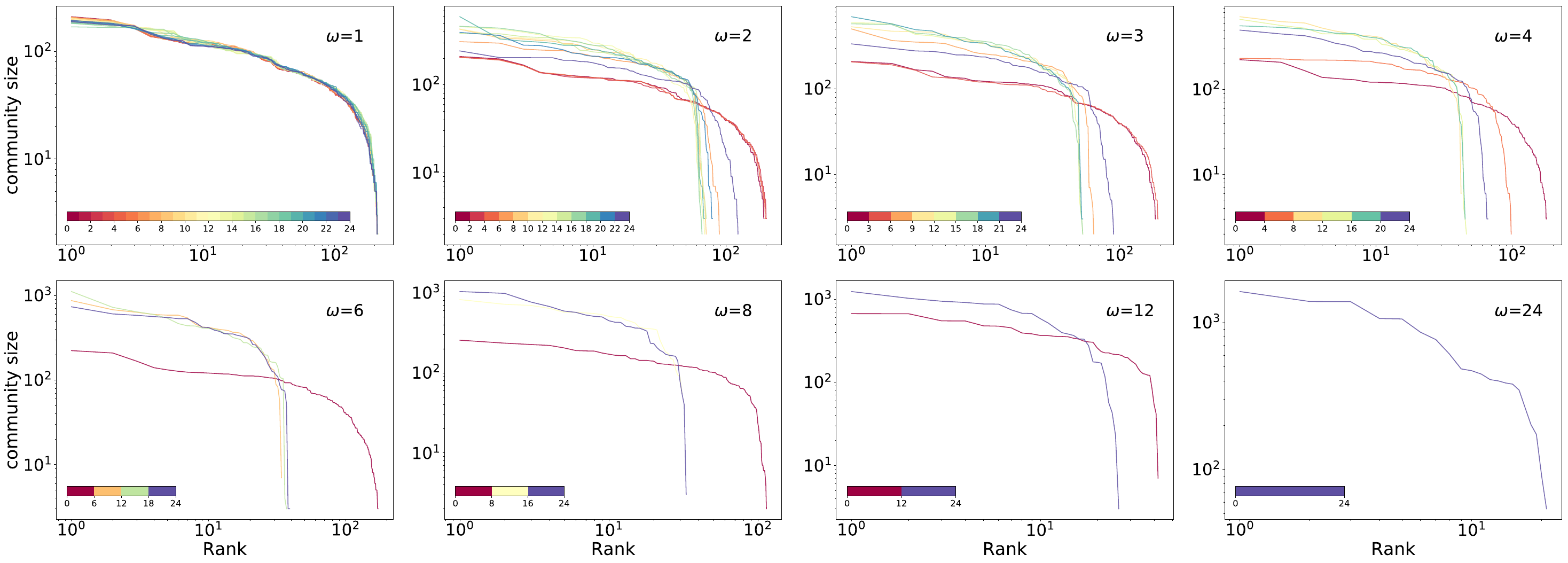}
\caption{\textbf{The Zipfian rank-size distribution of community size with varying time windows size $\omega$ in Abidjan.} Each line represents the distribution of a specific network in a certain time window. When $\omega=1$, all lines almost collapse together.  
The sleeping state of Abidjan is from 0 am to 6 am, which is clear from the figure with $\omega=2$, and the switching behavior always exists when $\omega\in[2,12]$.}
\label{fig:Abidjan}
\end{figure}

\begin{figure}[htb!] \centering
\includegraphics[width=1\linewidth]{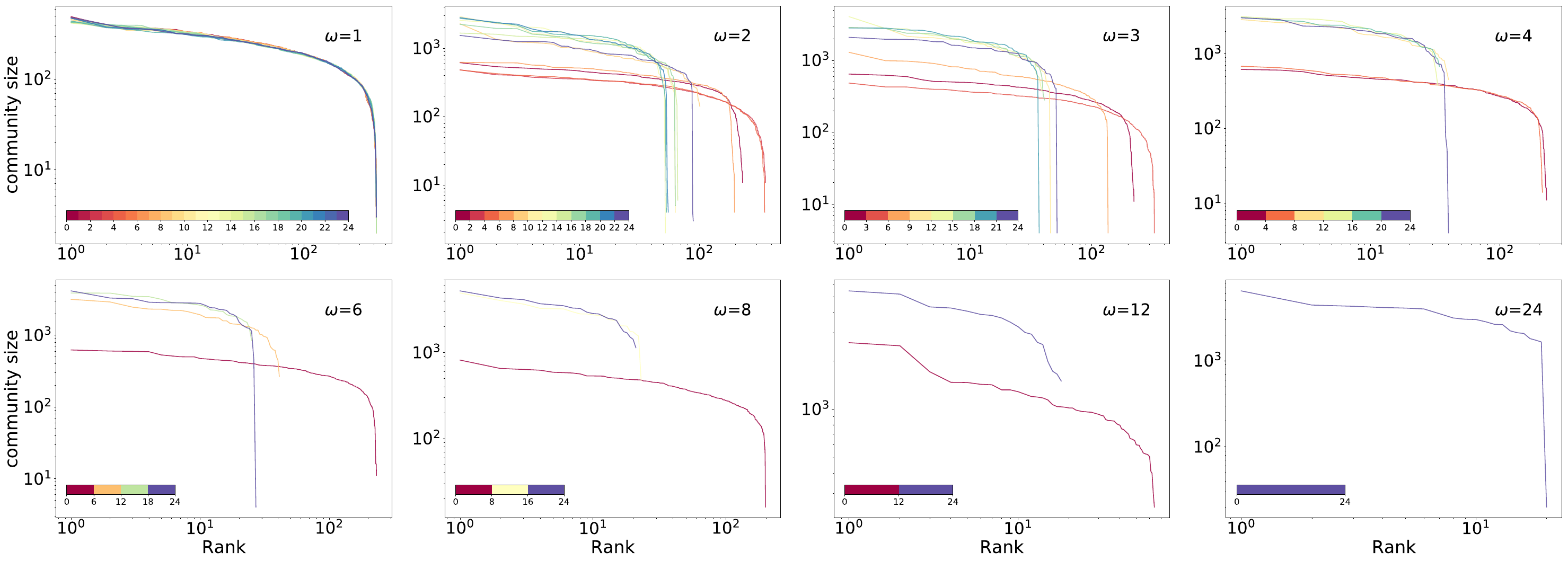}
\caption{\textbf{The Zipfian rank-size distribution of community size with varying time windows size $\omega$ in Dakar.} Each line represents the distribution of a specific network in a certain time window. When $\omega=1$, all lines almost collapse together.  
The sleeping state of Dakar is from 0 am to 8 am, which is clear from the figure with $\omega=2$, and the switching behavior always exists when $\omega\in[2,12]$.}
\label{fig:Dakar}
\end{figure}

When the time window size $\omega=1$ (see Supplementary Figs. \ref{fig:beijing}-\ref{fig:Dakar}), each identified community is just a clique in a specific location, and the community size is simply the number of individuals staying at that location at that time. Thus the size of the community can be interpreted as the hourly dynamic population of the location. Given the fact that people are almost constantly moving and hot-spots in cities are changing in a day (see Supplementary Fig. \ref{fig:hot_spots}), it is intriguing that the hourly dynamic population is stable over time in each city, which can be depicted by a truncated power-law (see Supplementary Fig. \ref{fig:fit}). The scaling exponent $\alpha$ is an urban characteristic and is larger in bigger cities. 

\begin{figure}[htb!] \centering
\includegraphics[width=0.8\linewidth]{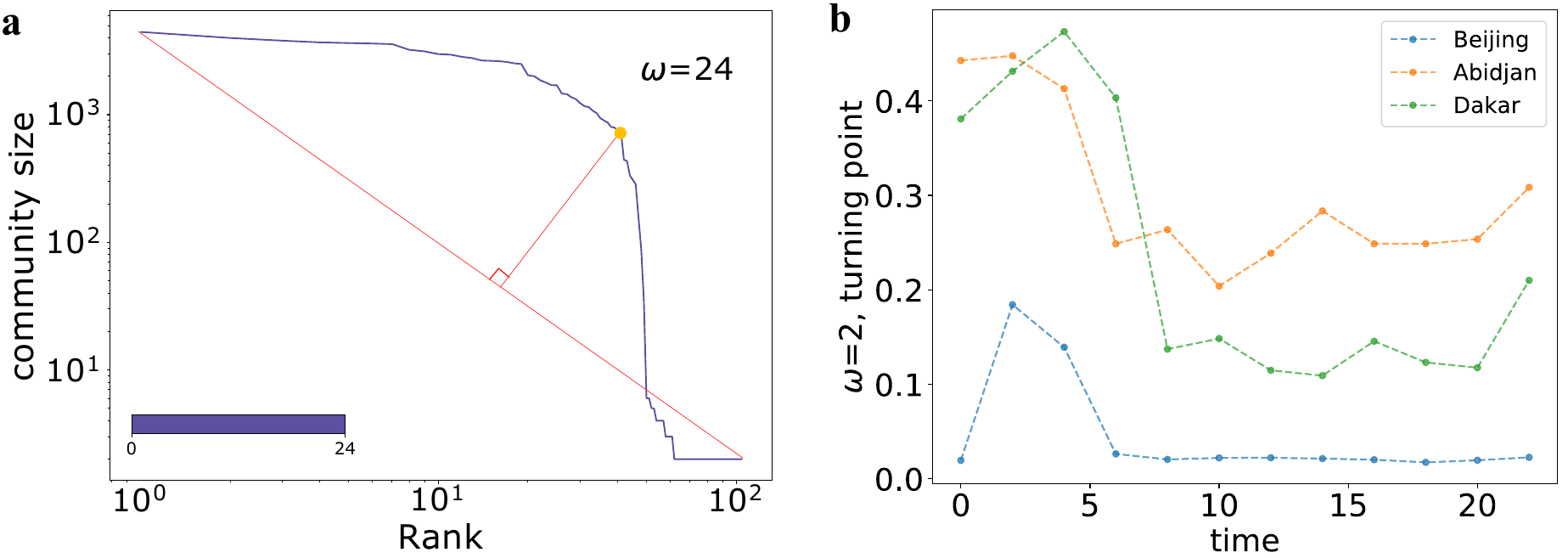}
\caption{\textbf{Detection of switching behavior between two interaction states via finding the turning point of distributions.}  \textbf{a} Detection of the turning point of distribution. We first connect the two ends of the curve (i.e. the first point and last point on the curve). Then the perpendicular distance is calculated between the straight line and every point of the curve. The point with the longest perpendicular distance is identified as the turning point. \textbf{b} the fractional rank value of the turning points of rank distributions of community size when $\omega=2$. For Beijing, turning points are of a larger value from 2 am to 6 am. In Abidjan, they are from 0 am to 6 am, and in Dakar, from 0 am to 8 am.}
\label{fig:curve}
\end{figure}

When aggregating spatio-temporal networks over consecutive time windows (i.e., when $\omega>1$), the switching behavior between active and sleeping states emerges. When $\omega=2$, the length of two states can be quantitatively identified via a geometric method, which finds the furthest point to the line that connected two ends of the curve (see Methods in the main text and Supplementary Fig. \ref{fig:curve}). For Beijing, when it is in the active state, the number of large-scale interaction communities ($>500$) is generally between 100 and 200. 
For Abidjan, this number is around 100. And for Dakar, it is less than 100. 

As the time window size $\omega$ increases from $2$ to $12$, the switching behavior between two interaction states remains, but the difference between distributions of two modes shrinks. The most prominent example is the cases when $\omega=12$. The intuition behind that is the ``inactive'' mode normally last from 4-8 hours, thus aggregating networks of larger time windows will mix the patterns of both sleeping and active periods, and active periods dominate the distribution. 
When $\omega=24$, Beijing is divided into 94 communities, in specific, 30 of these communities are larger than 1,000. Abidjan eventually has 110 communities. Due to a large number of individuals and the few towers in Dakar, only 20 communities were found, where 19 of them are larger than 1,000.


\section*{Supplementary Note 3: Robustness tests}

All results shown in Supplementary Figs. \ref{fig:beijing}-\ref{fig:Dakar} are obtained from one day's data. We use the data of the first day (December 1, 2013) in Beijing dataset, While use the the eighth day (December 13, 2011) in Abidjan dataset, and the second day (January 8, 2013) in Dakar dataset, as in these days there are more active users. We conduct the same data processing procedures and analysis in other days from the datasets in all three cities. All discoveries we presented in the main text well hold in other days (see Supplementary Figs. \ref{fig:beijing_days}-\ref{fig:Dakar_days}).

\begin{figure}[htb!] \centering
\includegraphics[width=1\linewidth]{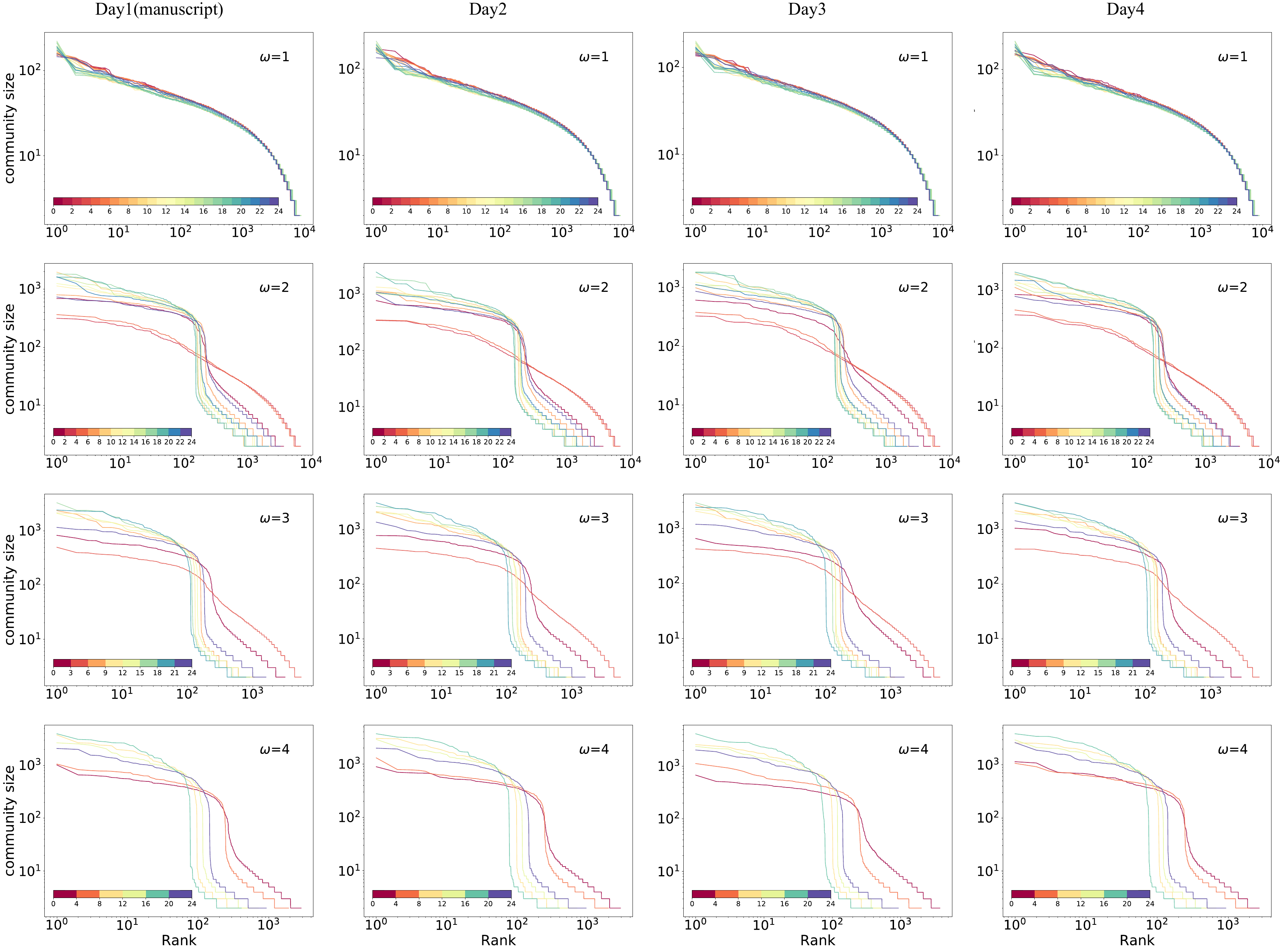}
\caption{\textbf{The Zipfian rank-size distribution of community size with varying time windows size $\omega$ in different days in Beijing.} Each line represents the distribution of a particular network on a different day at the same time window. Each column represents the distribution of a specific network on the same day in a specific time window, where the day1 is the Beijing data in the main text. The sleeping state of Beijing is from 2 am to 6 am, which is clear from the figure with $\omega=2$, and the switching behavior always exists when $\omega\in[2,4]$. It can be seen from the picture that the results of different days in Beijing are almost identical.}
\label{fig:beijing_days}
\end{figure}

\begin{figure}[htb!] \centering
\includegraphics[width=1\linewidth]{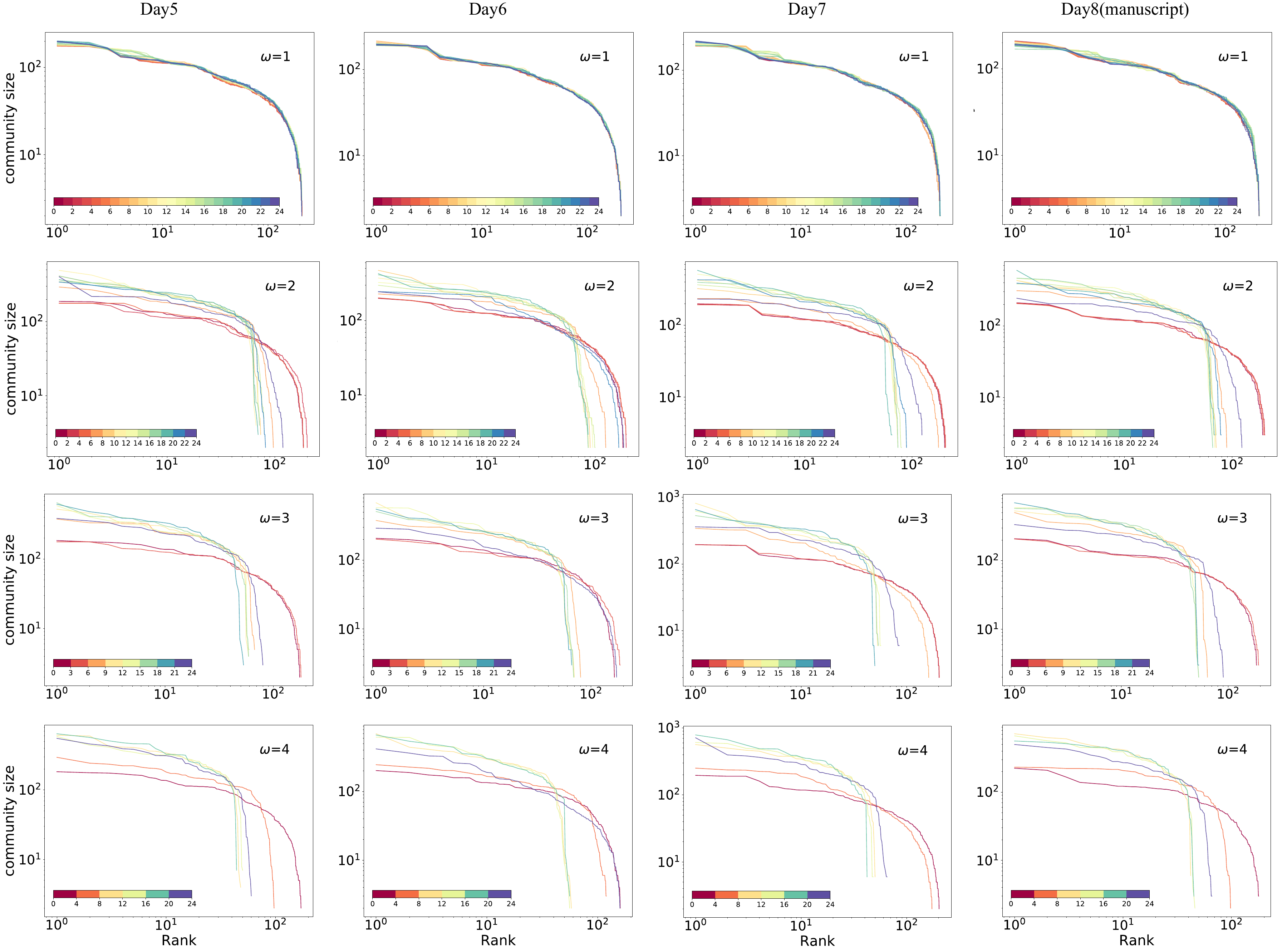}
\caption{\textbf{The Zipfian rank-size distribution of community size with varying time windows size $\omega$ in different days in Abidjan.} Each line represents the distribution of a particular network on a different day at the same time window. Each column represents the distribution of a specific network on the same day in a specific time window, where the day2 is the Beijing data in the main text. The sleeping state of Abidjan is from 0 am to 6 am, which is clear from the figure with $\omega=2$, and the switching behavior always exists when $\omega\in[2,4]$. It can be seen from the picture that the results of different days in Abidjan are almost identical.}
\label{fig:Abidjan_days}
\end{figure}

\begin{figure}[htb!] \centering
\includegraphics[width=1\linewidth]{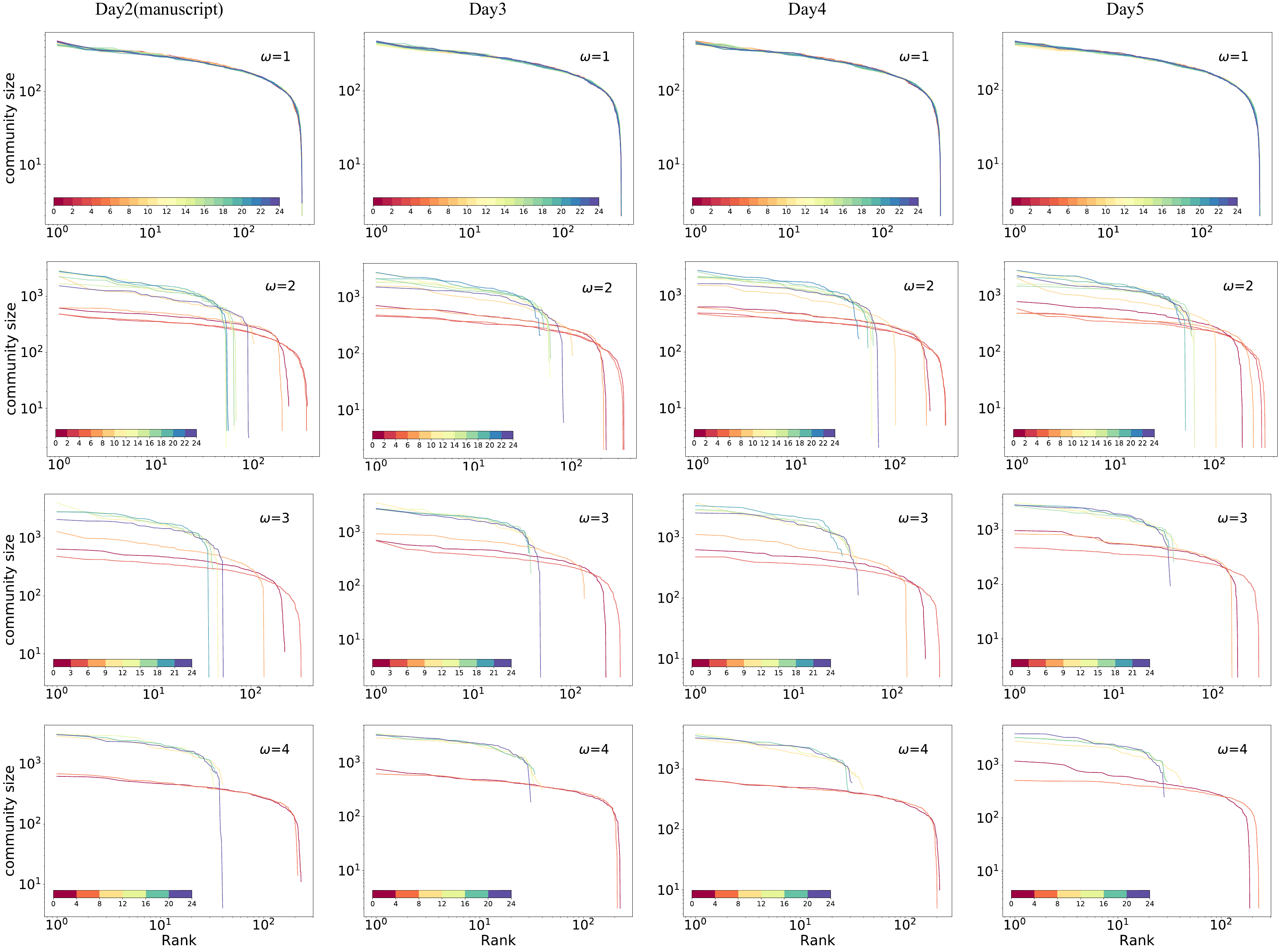}
\caption{\textbf{The Zipfian rank-size distribution of community size with varying time windows size $\omega$ in different days in Dakar.} Each line represents the distribution of a particular network on a different day at the same time window. Each column represents the distribution of a specific network on the same day in a specific time window, where the day8 is the Beijing data in the main text. The sleeping state of Dakar is from 0 am to 8 am, which is clear from the figure with $\omega=2$, and the switching behavior always exists when $\omega\in[2,4]$. It can be seen from the picture that the results of different days in Dakar are almost identical.}
\label{fig:Dakar_days}
\end{figure}


\begin{figure}[htb!] \centering
\includegraphics[width=1\linewidth]{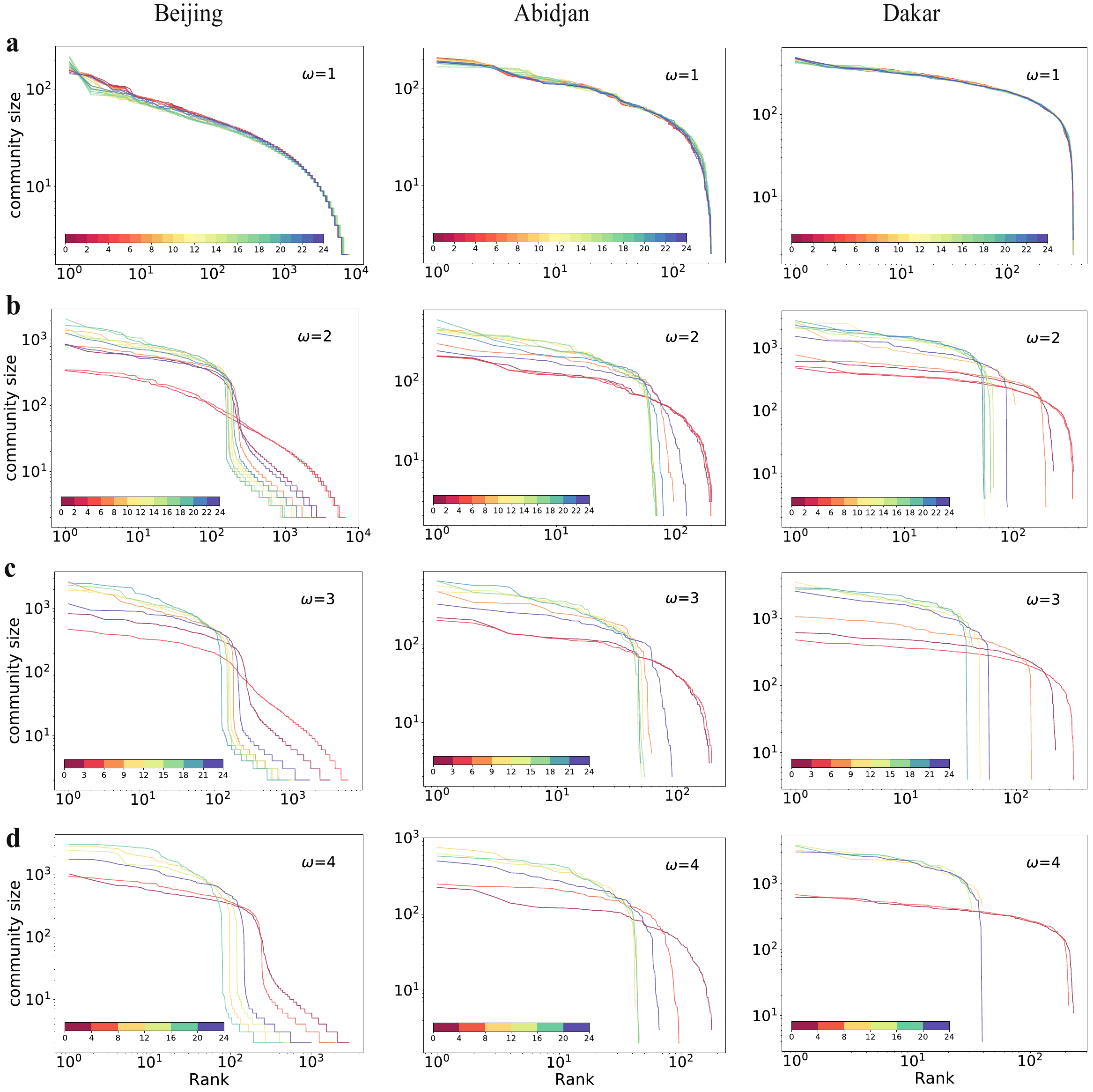}
\caption{\textbf{The Zipfian rank-size distribution of community size with varying time windows size $\omega$ for three cities when the interaction probability between individuals at the same location at the same time is 0.5.} Collective spatio-temporal interaction patterns remain almost the same as the Supplementary Figs. \ref{fig:beijing}-\ref{fig:Dakar}.}
\label{fig:interaction_prop}
\end{figure}

The interaction intensity or interaction probability (denote as $p$) depends on the type of the land. For example, in a traditional market or some event in a square, people would have a much higher probability to interact with each other; while in some other types of locations, e.g., office buildings, people in the same room might more likely to have fully interactions with each other and people in different rooms might have fewer interactions. In the main text, for simplicity and without losing generality, we assume the interaction probability $p$ equals 1. 
Here, we perform further tests with $p=0.5$, and we find that spatio-temporal interaction patterns reported in the main text remain even quantitatively the same (see Supplementary Fig. \ref{fig:interaction_prop}). The main difference is the number of spatio-temporal interaction links in the network, which is smaller than the network with $p=1$ but has a mild impact.


\begin{figure}[htb!] \centering
\includegraphics[width=1\linewidth]{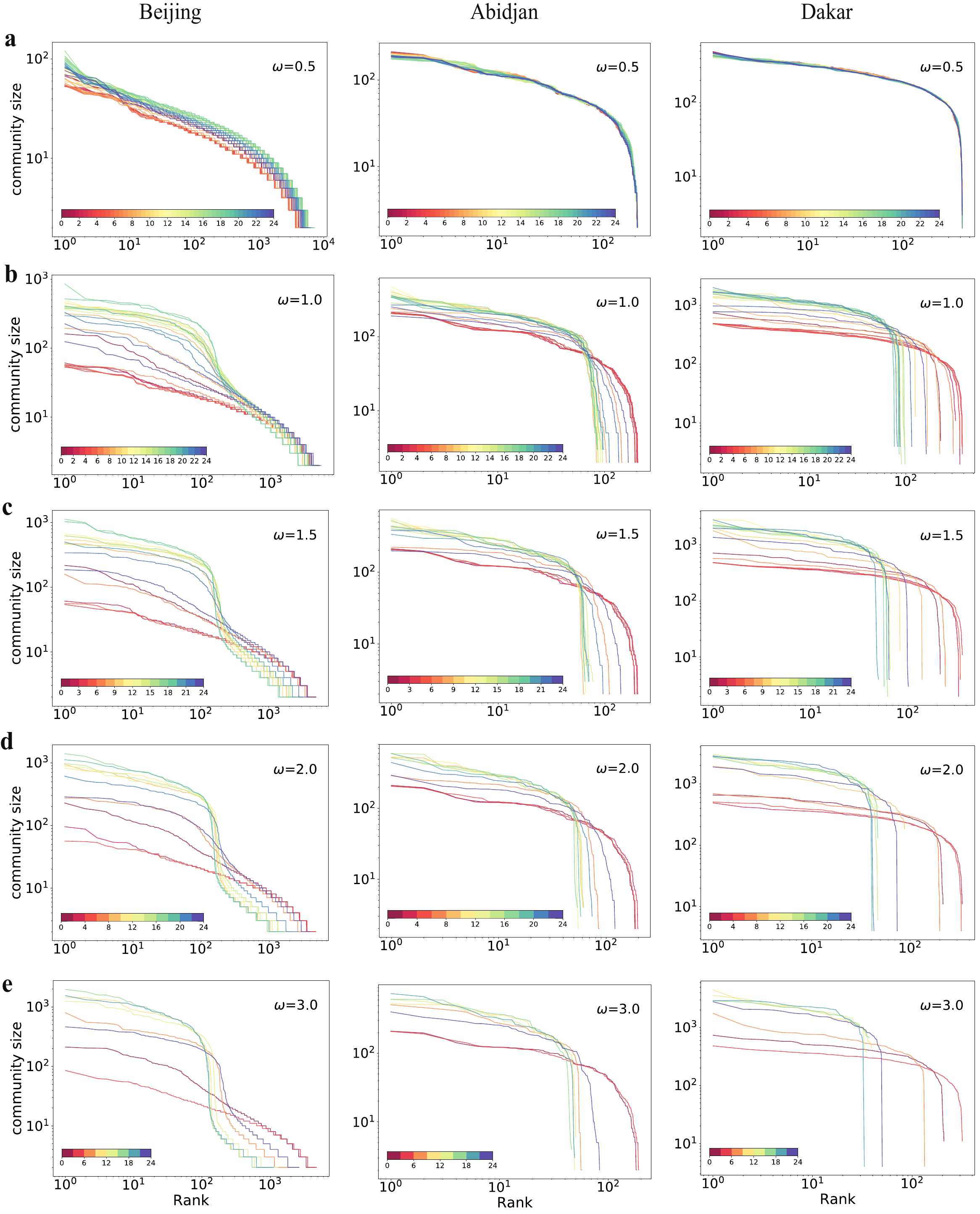}
\caption{\textbf{The Zipfian rank-size distribution of community size with varying time windows size $\omega$ for three cities when the unit time window is 0.5 hours.} Collective spatio-temporal interaction patterns remain almost the same as the Supplementary Fig. \ref{fig:beijing}-\ref{fig:Dakar}. }
\label{fig:time_window_30min}
\end{figure}

In the main text, the size of the unit time window is set as 1 hour to ensure that meaningful interactions can happen during it. We further test the impact of the unitary window size 
and set it as 0.5 hours, and we find that most results regarding spatio-temporal interaction patterns remain qualitatively same and quantitatively close (see Supplementary Fig. \ref{fig:time_window_30min}).

\section*{Supplementary Note 4: Comparisons between mobility models and impacts of different types of population}

To evaluate the impacts of heterogeneous departure probability on spatio-temporal interaction patterns, we employ the original PWO model \cite{yan2014universal} to make predictions for multiple steps as a null model. In the PWO model, at each time step, the probability of an individual moving from $i$ to $j$ is calculated from Eq. 2 in the main text. And the fraction of people who will move is the same at each time step (i.e., we assume $p^m(h)$ is uniform in this null model). The simulation results suggest that original PWO can not reproduce stable dynamical hourly population and switching behaviors (see Supplementary Figs. \ref{fig:beijing_model}-\ref{fig:Dakar_model}a,b) whereas, TPWO yields results that better reproduce the empirical observations (see Supplementary Fig. \ref{fig:beijing_model}-\ref{eq:PWO}-\ref{fig:Dakar_model}a,c). Therefore, the time-dependent departure probability plays an essential role in spatio-temporal interaction patterns at the urban scale. 

\begin{figure}[htb!] \centering
\includegraphics[width=1\linewidth]{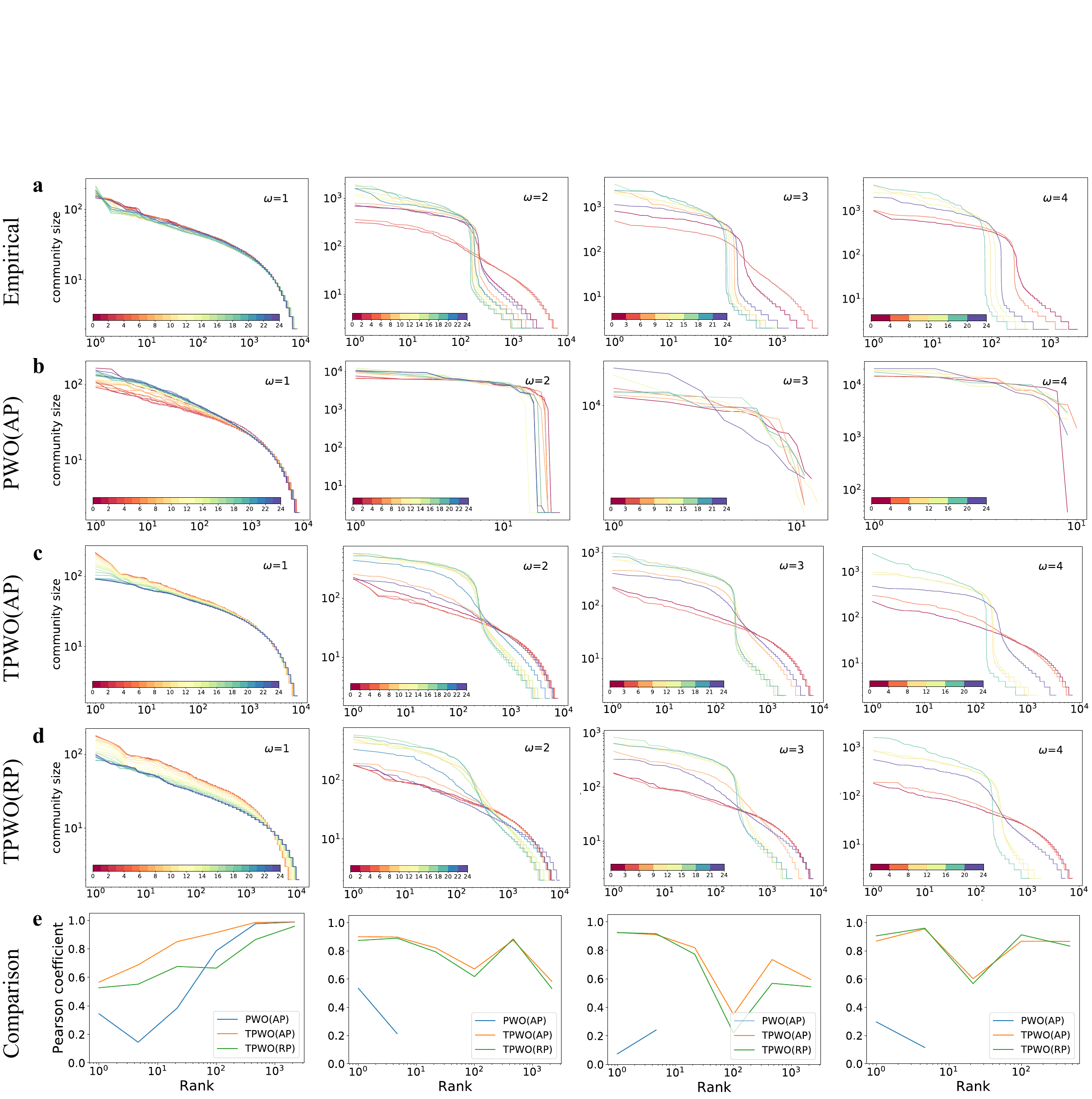}
\caption{\textbf{Comparisons between simulation resutls generated by PWO, TPWO(AP) and TPWO(RP) models with empirical obervations in Beijing.} \textbf{a} The empirical results of spatio-temporal interaction networks in Beijing. \textbf{b} Simulation results generated by PWO model without time-dependent departure probability (i.e., $p^m(h)$ is uniform). As for mobility models, active population (AP) has better performance than residential population (RP), we use AP for PWO model. \textbf{c,d} The simulation results generated by TPWO model with population $m_j$ of locations as AP (\textbf{c}) and RP (\textbf{d}). Choosing different type of population is important since it yields different results. \textbf{e} By dividing the community size distributions into intervals and comparing the Pearson coefficients between empirical observations simulation results, we find that TPWO(AP) always has a better performance.}
\label{fig:beijing_model}
\end{figure}

\begin{figure}[htb!] \centering
\includegraphics[width=1\linewidth]{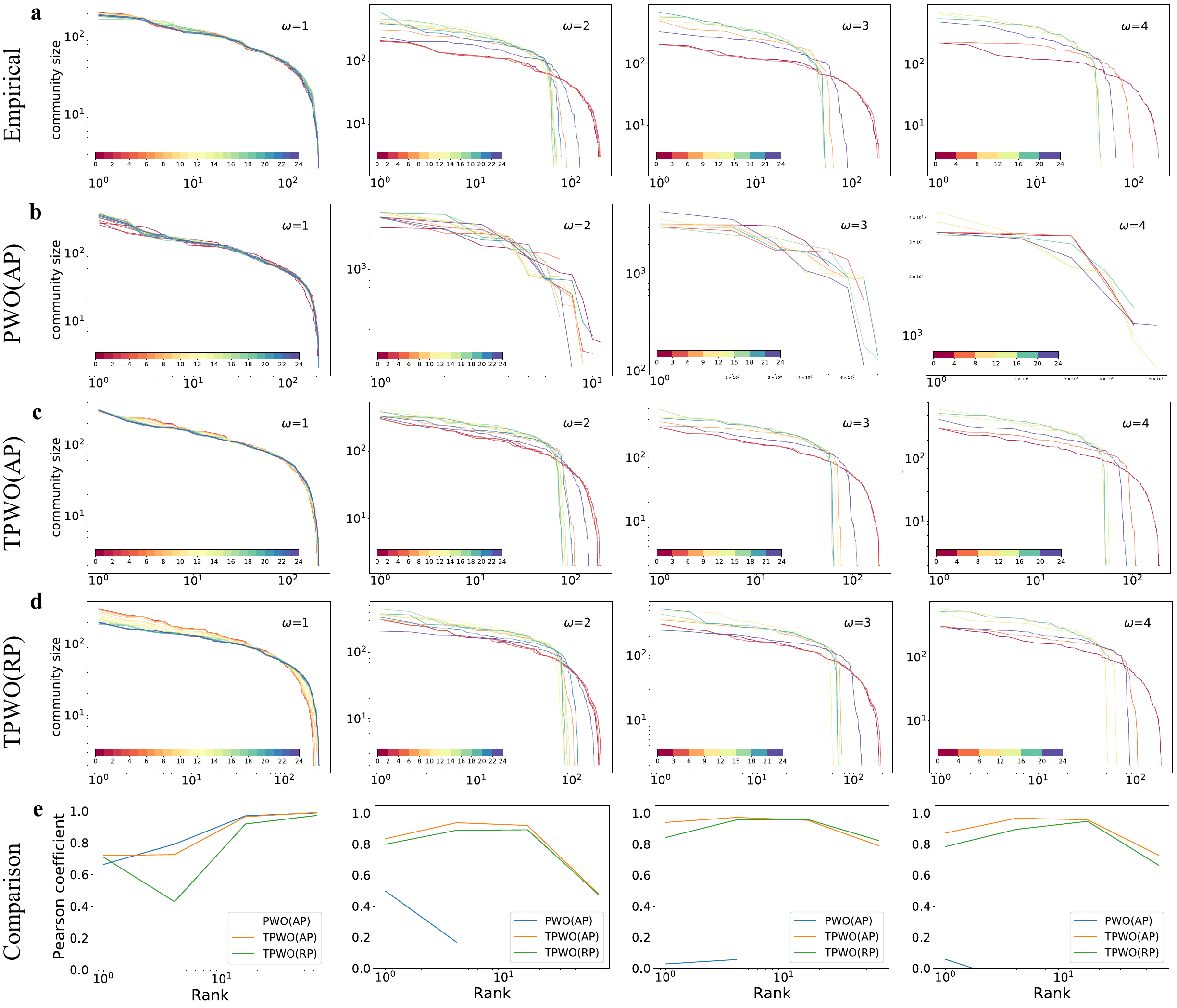}
\caption{\textbf{Comparisons between simulation resutls generated by PWO, TPWO(AP) and TPWO(RP) models with empirical obervations in Abidjan.} \textbf{a} The empirical results of spatio-temporal interaction networks in Abidjan. \textbf{b} Simulation results generated by PWO model without time-dependent departure probability (i.e., $p^m(h)$ is uniform). As for mobility models, active population (AP) has better performance than residential population (RP), we use AP for PWO model. \textbf{c,d} The simulation results generated by TPWO model with population $m_j$ of locations as AP (\textbf{c}) and RP (\textbf{d}). Choosing different type of population is important since it yields different results. \textbf{e} By dividing the community size distributions into intervals and comparing the Pearson coefficients between empirical observations simulation results, we find that TPWO(AP) always has a better performance. }
\label{fig:Abidjan_model}
\end{figure}

\newpage

\begin{figure}[htb!] \centering
\includegraphics[width=1\linewidth]{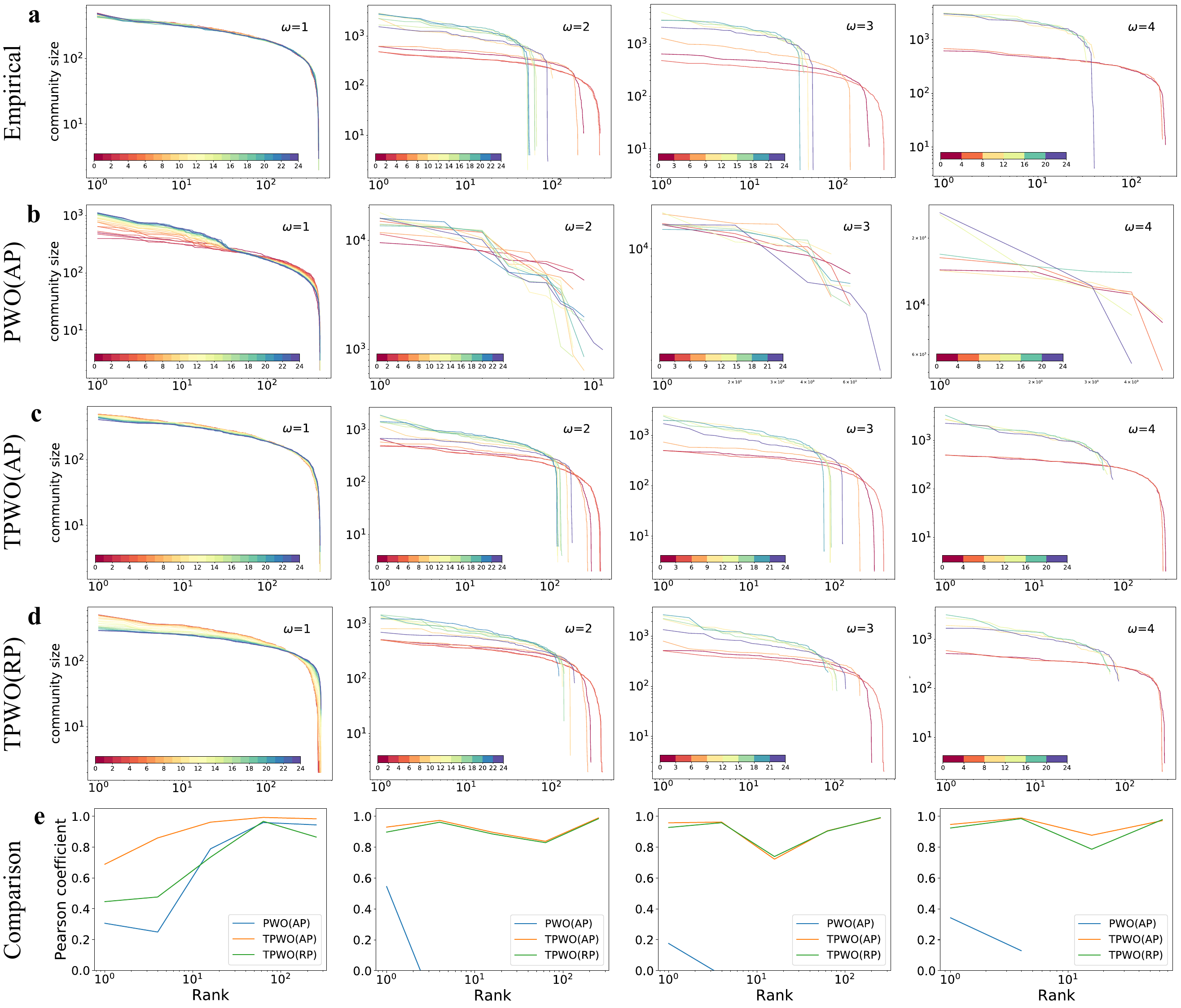}
\caption{\textbf{Comparisons between simulation resutls generated by PWO, TPWO(AP) and TPWO(RP) models with empirical obervations in Dakar.} \textbf{a} The empirical results of spatio-temporal interaction networks in Dakar. \textbf{b} Simulation results generated by PWO model without time-dependent departure probability (i.e., $p^m(h)$ is uniform). As for mobility models, active population (AP) has better performance than residential population (RP), we use AP for PWO model. \textbf{c,d} The simulation results generated by TPWO model with population $m_j$ of locations as AP (\textbf{c}) and RP (\textbf{d}). Choosing different type of population is important since it yields different results. \textbf{e} By dividing the community size distributions into intervals and comparing the Pearson coefficients between empirical observations simulation results, we find that TPWO(AP) always has a better performance.}
\label{fig:Dakar_model}
\end{figure}

The population of locations plays a critical role in mobility models. The most commonly used one is the residential population, but it is biased to CBD and working zones. Here, we test two versions of TPWO model with different choices of populations. One is the residential population (RP), for which we obtain it from WorldPop; the other one is the active population (AP) that depicts the number of people who ever visited the location in a day \cite{li2017simple}. AP can be estimated from massive cellphone data. 
The difference between results of two versions of TPWO is large especially when $\omega=1$, and TPWO(AP) model has better performance on explaining the behaviors of the stable hourly dynamic population (see $\omega=1$ cases in Supplementary Figs. \ref{fig:beijing_model}-\ref{fig:Dakar_model}c-e). When $\omega>1$, the impacts of different types of population decrease, but AP still has a better performance than RP. However, in many cases, AP is not easy to obtain when the access to cellphone data is limited. In the future, with better-developed sensor networks, AP might be more easily accessed. For example, Baidu now provides a heat map of the dynamic population in cities. 
Overall, using AP rather than RP yields better results for mobility models.

\begin{figure}[htb!] \centering
\includegraphics[width=1\linewidth]{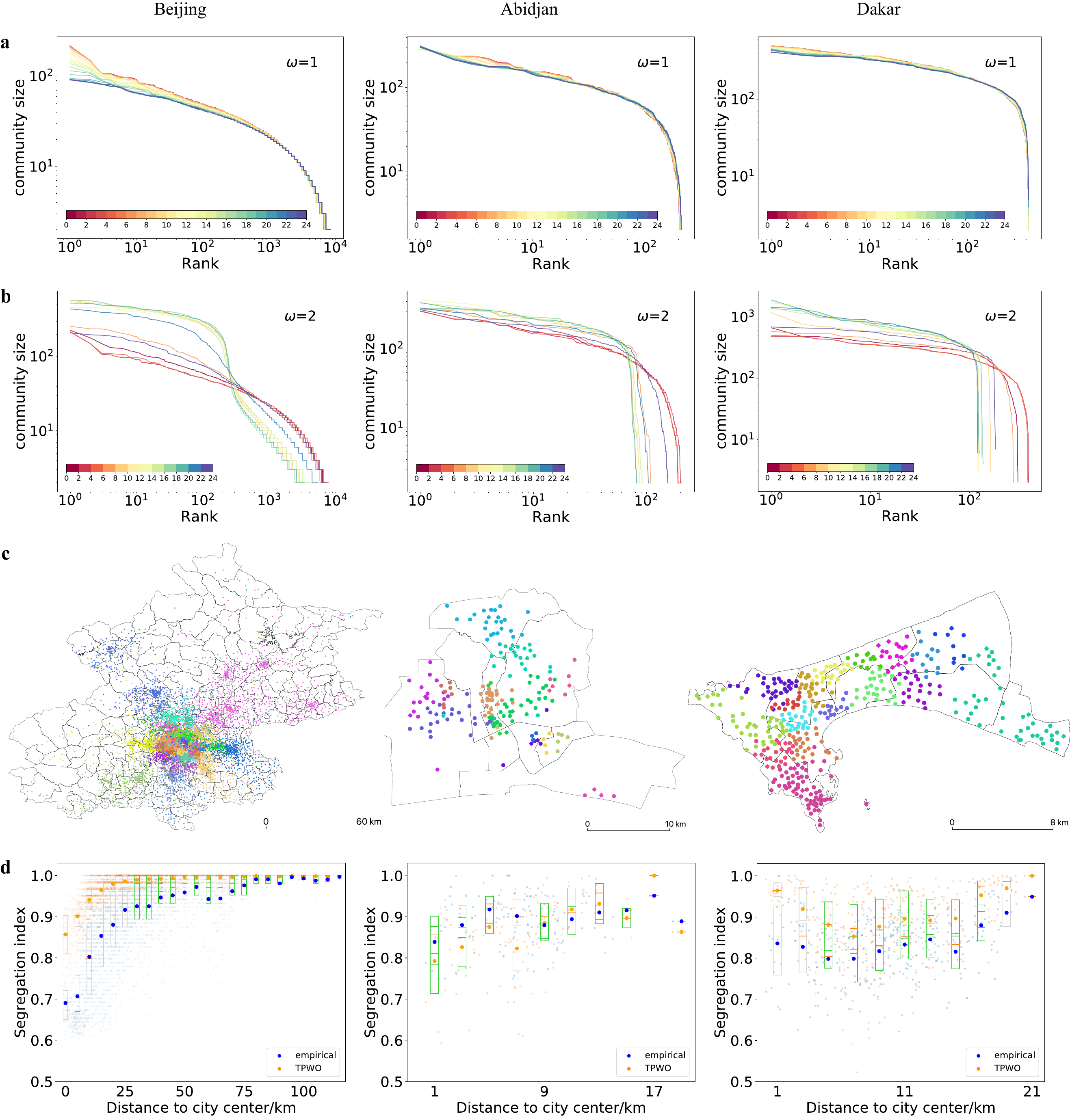}
\caption{\textbf{Spatio-temporal interaction patterns generated by the TPWO model with one trip per person per day on average, which is qualitatively the same with Fig. 5 in the main text. } \textbf{a} Rank-size distributions of communities in spatio-temporal networks constructed from results generated by the TPWO model, when $\omega=1$ and \textbf{b} $\omega=2$. \textbf{c} The spatial distribution of spatio-temporal interaction network communities generated by the TPWO model. \textbf{d} Quantitative comparisons between results generated by the TPWO model and the empirical ones from the city center to suburbs. The larger darker dots are the average, and smaller lighter dots represent results of each location. The rectangle indicates the 25\% and 75\% percentile of the data, and the darker bars in the rectangle are the median. When rectangles overlap, they are marked in green, otherwise, in grey.  } 
\label{fig:TPWOresultsSI}
\end{figure}

\section*{Supplementary Note 5: ``Folding Beijing'' in reality}
In a fiction novel ``Folding Beijing'' \cite{jingfang2015folding}, the author depicts a magic realism segregated urban space that there are three classes of people who are living in different urban spaces of Beijing, and having a different allowable duration of activities, living conditions, and even different sun-light duration. 
People from different classes are not allowed to interact with each other, which is certainly interaction segregation.  
In some sense, it is just a fantasy, but it reflects the fear of such a segregated city where spatio-temporal interactions are impossible between people from different groups. 
In this work, our results on spatio-temporal interaction segregation indicate that the reality is still promising, the urban centers are shared by people from different groups, and centers are aggregators of diversity (see Fig. 3 in the main text). 
%


\providecommand{\noopsort}[1]{}\providecommand{\singleletter}[1]{#1}%
\begin{thebibliography}{10}
\urlstyle{rm}
\expandafter\ifx\csname url\endcsname\relax
  \def\url#1{\texttt{#1}}\fi
\expandafter\ifx\csname urlprefix\endcsname\relax\def\urlprefix{URL }\fi
\expandafter\ifx\csname doiprefix\endcsname\relax\def\doiprefix{DOI: }\fi
\providecommand{\bibinfo}[2]{#2}
\providecommand{\eprint}[2][]{\url{#2}}

\bibitem{sim2015great}
\bibinfo{author}{Sim, A.}, \bibinfo{author}{Yaliraki, S.~N.},
  \bibinfo{author}{Barahona, M.} \& \bibinfo{author}{Stumpf, M.~P.}
\newblock \bibinfo{journal}{\bibinfo{title}{Great cities look small}}.
\newblock {\emph{\JournalTitle{Journal of the Royal Society Interface}}}
  \textbf{\bibinfo{volume}{12}}, \bibinfo{pages}{20150315}
  (\bibinfo{year}{2015}).

\bibitem{west2018scale}
\bibinfo{author}{West, G.}
\newblock \emph{\bibinfo{title}{Scale: The universal laws of life, growth, and
  death in organisms, cities, and companies}} (\bibinfo{publisher}{Penguin},
  \bibinfo{year}{2018}).

\bibitem{li2017simple}
\bibinfo{author}{Li, R.} \emph{et~al.}
\newblock \bibinfo{journal}{\bibinfo{title}{Simple spatial scaling rules behind
  complex cities}}.
\newblock {\emph{\JournalTitle{Nature Communications}}}
  \textbf{\bibinfo{volume}{8}}, \bibinfo{pages}{1--7} (\bibinfo{year}{2017}).

\bibitem{bettencourt2013origins}
\bibinfo{author}{Bettencourt, L.~M.}
\newblock \bibinfo{journal}{\bibinfo{title}{The origins of scaling in cities}}.
\newblock {\emph{\JournalTitle{Science}}} \textbf{\bibinfo{volume}{340}},
  \bibinfo{pages}{1438--1441} (\bibinfo{year}{2013}).

\bibitem{bettencourt2007growth}
\bibinfo{author}{Bettencourt, L.~M.}, \bibinfo{author}{Lobo, J.},
  \bibinfo{author}{Helbing, D.}, \bibinfo{author}{K{\"u}hnert, C.} \&
  \bibinfo{author}{West, G.~B.}
\newblock \bibinfo{journal}{\bibinfo{title}{Growth, innovation, scaling, and
  the pace of life in cities}}.
\newblock {\emph{\JournalTitle{Proceedings of the National Academy of
  Sciences}}} \textbf{\bibinfo{volume}{104}}, \bibinfo{pages}{7301--7306}
  (\bibinfo{year}{2007}).

\bibitem{alfeo2019assessing}
\bibinfo{author}{Alfeo, A.~L.}, \bibinfo{author}{Cimino, M.~G.},
  \bibinfo{author}{Lepri, B.}, \bibinfo{author}{Pentland, A.~S.} \&
  \bibinfo{author}{Vaglini, G.}
\newblock \bibinfo{journal}{\bibinfo{title}{Assessing refugees’ integration
  via spatio-temporal similarities of mobility and calling behaviors}}.
\newblock {\emph{\JournalTitle{IEEE Transactions on Computational Social
  Systems}}} \textbf{\bibinfo{volume}{6}}, \bibinfo{pages}{726--738}
  (\bibinfo{year}{2019}).

\bibitem{logan1987racial}
\bibinfo{author}{Logan, J.~R.} \& \bibinfo{author}{Messner, S.~F.}
\newblock \bibinfo{journal}{\bibinfo{title}{Racial residential segregation and
  suburban violent crime}}.
\newblock {\emph{\JournalTitle{Social Science Quarterly}}}
  \textbf{\bibinfo{volume}{68}}, \bibinfo{pages}{510} (\bibinfo{year}{1987}).

\bibitem{acevedo2003residential}
\bibinfo{author}{Acevedo-Garcia, D.} \& \bibinfo{author}{Lochner, K.~A.}
\newblock \bibinfo{journal}{\bibinfo{title}{Residential segregation and
  health}}.
\newblock {\emph{\JournalTitle{Neighborhoods and Health}}}
  \bibinfo{pages}{265--87} (\bibinfo{year}{2003}).

\bibitem{toth2021inequality}
\bibinfo{author}{T{\'o}th, G.} \emph{et~al.}
\newblock \bibinfo{journal}{\bibinfo{title}{Inequality is rising where social
  network segregation interacts with urban topology}}.
\newblock {\emph{\JournalTitle{Nature Communications}}}
  \textbf{\bibinfo{volume}{12}}, \bibinfo{pages}{1--9} (\bibinfo{year}{2021}).

\bibitem{jacobs2016death}
\bibinfo{author}{Jacobs, J.}
\newblock \emph{\bibinfo{title}{The death and life of great American cities}}
  (\bibinfo{publisher}{Vintage}, \bibinfo{year}{2016}).

\bibitem{song2010modelling}
\bibinfo{author}{Song, C.}, \bibinfo{author}{Koren, T.}, \bibinfo{author}{Wang,
  P.} \& \bibinfo{author}{Barab{\'a}si, A.-L.}
\newblock \bibinfo{journal}{\bibinfo{title}{Modelling the scaling properties of
  human mobility}}.
\newblock {\emph{\JournalTitle{Nature Physics}}} \textbf{\bibinfo{volume}{6}},
  \bibinfo{pages}{818--823} (\bibinfo{year}{2010}).

\bibitem{schlapfer2021universal}
\bibinfo{author}{Schl{\"a}pfer, M.} \emph{et~al.}
\newblock \bibinfo{journal}{\bibinfo{title}{The universal visitation law of
  human mobility}}.
\newblock {\emph{\JournalTitle{Nature}}} \textbf{\bibinfo{volume}{593}},
  \bibinfo{pages}{522--527} (\bibinfo{year}{2021}).

\bibitem{alessandretti2020scales}
\bibinfo{author}{Alessandretti, L.}, \bibinfo{author}{Aslak, U.} \&
  \bibinfo{author}{Lehmann, S.}
\newblock \bibinfo{journal}{\bibinfo{title}{The scales of human mobility}}.
\newblock {\emph{\JournalTitle{Nature}}} \textbf{\bibinfo{volume}{587}},
  \bibinfo{pages}{402--407} (\bibinfo{year}{2020}).

\bibitem{pan2013urban}
\bibinfo{author}{Pan, W.}, \bibinfo{author}{Ghoshal, G.},
  \bibinfo{author}{Krumme, C.}, \bibinfo{author}{Cebrian, M.} \&
  \bibinfo{author}{Pentland, A.}
\newblock \bibinfo{journal}{\bibinfo{title}{Urban characteristics attributable
  to density-driven tie formation}}.
\newblock {\emph{\JournalTitle{Nature Communications}}}
  \textbf{\bibinfo{volume}{4}}, \bibinfo{pages}{1--7} (\bibinfo{year}{2013}).

\bibitem{smith1776wealth}
\bibinfo{author}{Smith, A.}
\newblock \emph{\bibinfo{title}{The wealth of nations (1776)}}
  (\bibinfo{publisher}{New York: Modern Library}, \bibinfo{year}{1937}).

\bibitem{balcan2009multiscale}
\bibinfo{author}{Balcan, D.} \emph{et~al.}
\newblock \bibinfo{journal}{\bibinfo{title}{Multiscale mobility networks and
  the spatial spreading of infectious diseases}}.
\newblock {\emph{\JournalTitle{Proceedings of the National Academy of
  Sciences}}} \textbf{\bibinfo{volume}{106}}, \bibinfo{pages}{21484--21489}
  (\bibinfo{year}{2009}).

\bibitem{schlapfer2014scaling}
\bibinfo{author}{Schl{\"a}pfer, M.} \emph{et~al.}
\newblock \bibinfo{journal}{\bibinfo{title}{The scaling of human interactions
  with city size}}.
\newblock {\emph{\JournalTitle{Journal of the Royal Society Interface}}}
  \textbf{\bibinfo{volume}{11}}, \bibinfo{pages}{20130789}
  (\bibinfo{year}{2014}).

\bibitem{li2017effects}
\bibinfo{author}{Li, R.}, \bibinfo{author}{Wang, W.} \& \bibinfo{author}{Di,
  Z.}
\newblock \bibinfo{journal}{\bibinfo{title}{Effects of human dynamics on
  epidemic spreading in c{\^o}te d’ivoire}}.
\newblock {\emph{\JournalTitle{Physica A: Statistical Mechanics and its
  Applications}}} \textbf{\bibinfo{volume}{467}}, \bibinfo{pages}{30--40}
  (\bibinfo{year}{2017}).

\bibitem{deville2016scaling}
\bibinfo{author}{Deville, P.} \emph{et~al.}
\newblock \bibinfo{journal}{\bibinfo{title}{Scaling identity connects human
  mobility and social interactions}}.
\newblock {\emph{\JournalTitle{Proceedings of the National Academy of
  Sciences}}} \textbf{\bibinfo{volume}{113}}, \bibinfo{pages}{7047--7052}
  (\bibinfo{year}{2016}).

\bibitem{li2020early}
\bibinfo{author}{Li, Q.} \emph{et~al.}
\newblock \bibinfo{journal}{\bibinfo{title}{Early transmission dynamics in
  wuhan, china, of novel coronavirus--infected pneumonia}}.
\newblock {\emph{\JournalTitle{New England Journal of Medicine}}}
  (\bibinfo{year}{2020}).

\bibitem{oliver2020mobile}
\bibinfo{author}{Oliver, N.} \emph{et~al.}
\newblock \bibinfo{journal}{\bibinfo{title}{Mobile phone data for informing
  public health actions across the covid-19 pandemic life cycle}}.
\newblock {\emph{\JournalTitle{Science Advances}}}
  \textbf{\bibinfo{volume}{6}}, \bibinfo{pages}{eabc0764}
  (\bibinfo{year}{2020}).

\bibitem{li2018effect}
\bibinfo{author}{Li, R.}, \bibinfo{author}{Richmond, P.} \&
  \bibinfo{author}{Roehner, B.~M.}
\newblock \bibinfo{journal}{\bibinfo{title}{Effect of population density on
  epidemics}}.
\newblock {\emph{\JournalTitle{Physica A: Statistical Mechanics and its
  Applications}}} \textbf{\bibinfo{volume}{510}}, \bibinfo{pages}{713--724}
  (\bibinfo{year}{2018}).

\bibitem{zhong2021country}
\bibinfo{author}{Zhong, L.}, \bibinfo{author}{Diagne, M.},
  \bibinfo{author}{Wang, W.} \& \bibinfo{author}{Gao, J.}
\newblock \bibinfo{journal}{\bibinfo{title}{Country distancing increase reveals
  the effectiveness of travel restrictions in stopping covid-19 transmission}}.
\newblock {\emph{\JournalTitle{Communications Physics}}}
  \textbf{\bibinfo{volume}{4}}, \bibinfo{pages}{1--12} (\bibinfo{year}{2021}).

\bibitem{mo2021modeling}
\bibinfo{author}{Mo, B.} \emph{et~al.}
\newblock \bibinfo{journal}{\bibinfo{title}{Modeling epidemic spreading through
  public transit using time-varying encounter network}}.
\newblock {\emph{\JournalTitle{Transportation Research Part C: Emerging
  Technologies}}} \textbf{\bibinfo{volume}{122}}, \bibinfo{pages}{102893}
  (\bibinfo{year}{2021}).

\bibitem{batty2013new}
\bibinfo{author}{Batty, M.}
\newblock \emph{\bibinfo{title}{The new science of cities}}
  (\bibinfo{publisher}{MIT press}, \bibinfo{year}{2013}).

\bibitem{louf2016patterns}
\bibinfo{author}{Louf, R.} \& \bibinfo{author}{Barthelemy, M.}
\newblock \bibinfo{journal}{\bibinfo{title}{Patterns of residential
  segregation}}.
\newblock {\emph{\JournalTitle{PLoS One}}} \textbf{\bibinfo{volume}{11}},
  \bibinfo{pages}{e0157476} (\bibinfo{year}{2016}).

\bibitem{chodrow2017structure}
\bibinfo{author}{Chodrow, P.~S.}
\newblock \bibinfo{journal}{\bibinfo{title}{Structure and information in
  spatial segregation}}.
\newblock {\emph{\JournalTitle{Proceedings of the National Academy of
  Sciences}}} \textbf{\bibinfo{volume}{114}}, \bibinfo{pages}{11591--11596}
  (\bibinfo{year}{2017}).

\bibitem{wang2016daily}
\bibinfo{author}{Wang, D.} \& \bibinfo{author}{Li, F.}
\newblock \bibinfo{journal}{\bibinfo{title}{Daily activity space and exposure:
  A comparative study of hong kong's public and private housing residents'
  segregation in daily life}}.
\newblock {\emph{\JournalTitle{Cities}}} \textbf{\bibinfo{volume}{59}},
  \bibinfo{pages}{148--155} (\bibinfo{year}{2016}).

\bibitem{xu2019quantifying}
\bibinfo{author}{Xu, Y.}, \bibinfo{author}{Belyi, A.}, \bibinfo{author}{Santi,
  P.} \& \bibinfo{author}{Ratti, C.}
\newblock \bibinfo{journal}{\bibinfo{title}{Quantifying segregation in an
  integrated urban physical-social space}}.
\newblock {\emph{\JournalTitle{Journal of the Royal Society Interface}}}
  \textbf{\bibinfo{volume}{16}}, \bibinfo{pages}{20190536}
  (\bibinfo{year}{2019}).

\bibitem{helbing2001traffic}
\bibinfo{author}{Helbing, D.}
\newblock \bibinfo{journal}{\bibinfo{title}{Traffic and related self-driven
  many-particle systems}}.
\newblock {\emph{\JournalTitle{Reviews of Modern Physics}}}
  \textbf{\bibinfo{volume}{73}}, \bibinfo{pages}{1067} (\bibinfo{year}{2001}).

\bibitem{de2011modelling}
\bibinfo{author}{de~Dios~Ort{\'u}zar, J.} \& \bibinfo{author}{Willumsen, L.~G.}
\newblock \emph{\bibinfo{title}{Modelling transport}} (\bibinfo{publisher}{John
  wiley \& sons}, \bibinfo{year}{2011}).

\bibitem{li2021gravity}
\bibinfo{author}{Li, R.} \emph{et~al.}
\newblock \bibinfo{journal}{\bibinfo{title}{Gravity model in dockless
  bike-sharing systems within cities}}.
\newblock {\emph{\JournalTitle{Physical Review E}}}
  \textbf{\bibinfo{volume}{103}}, \bibinfo{pages}{012312}
  (\bibinfo{year}{2021}).

\bibitem{qiu2022understanding}
\bibinfo{author}{Qiu, X.} \emph{et~al.}
\newblock \bibinfo{journal}{\bibinfo{title}{Understanding urban congestion with
  biking traffic and routing detour ratio}}.
\newblock {\emph{\JournalTitle{arXiv preprint arXiv:2205.08118}}}
  (\bibinfo{year}{2022}).

\bibitem{li2022emergence}
\bibinfo{author}{Li, R.} \emph{et~al.}
\newblock \bibinfo{journal}{\bibinfo{title}{Emergence of scaling in dockless
  bike-sharing systems}}.
\newblock {\emph{\JournalTitle{arXiv}}} \textbf{\bibinfo{volume}{2202.06352}}
  (\bibinfo{year}{2022}).

\bibitem{ruan2020dynamic}
\bibinfo{author}{Ruan, S.} \emph{et~al.}
\newblock \bibinfo{title}{Dynamic public resource allocation based on human
  mobility prediction}.
\newblock In \emph{\bibinfo{booktitle}{Proceedings of the ACM on Interactive,
  Mobile, Wearable and Ubiquitous Technologies}}, \bibinfo{number}{4(1)},
  \bibinfo{pages}{1--22} (\bibinfo{publisher}{ACM New York, NY, USA},
  \bibinfo{year}{2020}).

\bibitem{lu2012predictability}
\bibinfo{author}{Lu, X.}, \bibinfo{author}{Bengtsson, L.} \&
  \bibinfo{author}{Holme, P.}
\newblock \bibinfo{journal}{\bibinfo{title}{Predictability of population
  displacement after the 2010 haiti earthquake}}.
\newblock {\emph{\JournalTitle{Proceedings of the National Academy of
  Sciences}}} \textbf{\bibinfo{volume}{109}}, \bibinfo{pages}{11576--11581}
  (\bibinfo{year}{2012}).

\bibitem{bagrow2011collective}
\bibinfo{author}{Bagrow, J.~P.}, \bibinfo{author}{Wang, D.} \&
  \bibinfo{author}{Barabasi, A.-L.}
\newblock \bibinfo{journal}{\bibinfo{title}{Collective response of human
  populations to large-scale emergencies}}.
\newblock {\emph{\JournalTitle{PLoS One}}} \textbf{\bibinfo{volume}{6}},
  \bibinfo{pages}{e17680} (\bibinfo{year}{2011}).

\bibitem{eagle2006reality}
\bibinfo{author}{Eagle, N.} \& \bibinfo{author}{Pentland, A.~S.}
\newblock \bibinfo{journal}{\bibinfo{title}{Reality mining: sensing complex
  social systems}}.
\newblock {\emph{\JournalTitle{Personal and Ubiquitous Computing}}}
  \textbf{\bibinfo{volume}{10}}, \bibinfo{pages}{255--268}
  (\bibinfo{year}{2006}).

\bibitem{blondel2015survey}
\bibinfo{author}{Blondel, V.~D.}, \bibinfo{author}{Decuyper, A.} \&
  \bibinfo{author}{Krings, G.}
\newblock \bibinfo{journal}{\bibinfo{title}{A survey of results on mobile phone
  datasets analysis}}.
\newblock {\emph{\JournalTitle{EPJ Data Science}}}
  \textbf{\bibinfo{volume}{4}}, \bibinfo{pages}{10} (\bibinfo{year}{2015}).

\bibitem{xu2017clearer}
\bibinfo{author}{Xu, Y.}, \bibinfo{author}{Li, R.}, \bibinfo{author}{Jiang,
  S.}, \bibinfo{author}{Zhang, J.} \& \bibinfo{author}{Gonz{\'a}lez, M.~C.}
\newblock \bibinfo{title}{Clearer skies in beijing--revealing the impacts of
  traffic on the modeling of air quality}.
\newblock In \emph{\bibinfo{booktitle}{Proceedings of the Transportation
  Research Board (TRB) 96th Annual Meeting}}, \bibinfo{number}{17: 05211}
  (\bibinfo{year}{2017}).

\bibitem{xu2019unravel}
\bibinfo{author}{Xu, Y.} \emph{et~al.}
\newblock \bibinfo{journal}{\bibinfo{title}{Unravel the landscape and pulses of
  cycling activities from a dockless bike-sharing system}}.
\newblock {\emph{\JournalTitle{Computers, Environment and Urban Systems}}}
  \textbf{\bibinfo{volume}{75}}, \bibinfo{pages}{184--203}
  (\bibinfo{year}{2019}).

\bibitem{dong2016population}
\bibinfo{author}{Dong, L.}, \bibinfo{author}{Li, R.}, \bibinfo{author}{Zhang,
  J.} \& \bibinfo{author}{Di, Z.}
\newblock \bibinfo{journal}{\bibinfo{title}{Population-weighted efficiency in
  transportation networks}}.
\newblock {\emph{\JournalTitle{Scientific Reports}}}
  \textbf{\bibinfo{volume}{6}}, \bibinfo{pages}{1--10} (\bibinfo{year}{2016}).

\bibitem{blondel2012d4d}
\bibinfo{author}{Blondel, V.~D.} \emph{et~al.}
\newblock \bibinfo{journal}{\bibinfo{title}{Data for development: the d4d
  challenge on mobile phone data}}.
\newblock {\emph{\JournalTitle{ArXiv Preprint ArXiv:1210.0137}}}
  (\bibinfo{year}{2012}).

\bibitem{de2014d4d}
\bibinfo{author}{de~Montjoye, Y.-A.}, \bibinfo{author}{Smoreda, Z.},
  \bibinfo{author}{Trinquart, R.}, \bibinfo{author}{Ziemlicki, C.} \&
  \bibinfo{author}{Blondel, V.~D.}
\newblock \bibinfo{journal}{\bibinfo{title}{D4d-senegal: the second mobile
  phone data for development challenge}}.
\newblock {\emph{\JournalTitle{ArXiv Preprint ArXiv:1407.4885}}}
  (\bibinfo{year}{2014}).

\bibitem{alexander2015origin}
\bibinfo{author}{Alexander, L.}, \bibinfo{author}{Jiang, S.},
  \bibinfo{author}{Murga, M.} \& \bibinfo{author}{Gonz{\'a}lez, M.~C.}
\newblock \bibinfo{journal}{\bibinfo{title}{Origin--destination trips by
  purpose and time of day inferred from mobile phone data}}.
\newblock {\emph{\JournalTitle{Transportation Research Part C: Emerging
  Technologies}}} \textbf{\bibinfo{volume}{58}}, \bibinfo{pages}{240--250}
  (\bibinfo{year}{2015}).

\bibitem{ccolak2015analyzing}
\bibinfo{author}{{\c{C}}olak, S.}, \bibinfo{author}{Alexander, L.~P.},
  \bibinfo{author}{Alvim, B.~G.}, \bibinfo{author}{Mehndiratta, S.~R.} \&
  \bibinfo{author}{Gonz{\'a}lez, M.~C.}
\newblock \bibinfo{journal}{\bibinfo{title}{Analyzing cell phone location data
  for urban travel: current methods, limitations, and opportunities}}.
\newblock {\emph{\JournalTitle{Transportation Research Record}}}
  \textbf{\bibinfo{volume}{2526}}, \bibinfo{pages}{126--135}
  (\bibinfo{year}{2015}).

\bibitem{STcompanions}
\bibinfo{title}{Lockdown voices: Big-data glitches, yellow health-codes, and
  “spatial-temporal companions”}.
\newblock
  \bibinfo{howpublished}{\url{https://chinadigitaltimes.net%/2021/11/lockdown-voices-big-data-glitches-yellow-health-codes-and-spatial-temporal-companions
  }}.
\newblock \bibinfo{note}{Accessed: 2021-12-06}.

\bibitem{nie2021understanding}
\bibinfo{author}{Nie, W.-P.}, \bibinfo{author}{Zhao, Z.-D.},
  \bibinfo{author}{Cai, S.-M.} \& \bibinfo{author}{Zhou, T.}
\newblock \bibinfo{journal}{\bibinfo{title}{Understanding the urban mobility
  community by taxi travel trajectory}}.
\newblock {\emph{\JournalTitle{Communications in Nonlinear Science and
  Numerical Simulation}}} \textbf{\bibinfo{volume}{101}},
  \bibinfo{pages}{105863} (\bibinfo{year}{2021}).

\bibitem{blondel2008fast}
\bibinfo{author}{Blondel, V.~D.}, \bibinfo{author}{Guillaume, J.-L.},
  \bibinfo{author}{Lambiotte, R.} \& \bibinfo{author}{Lefebvre, E.}
\newblock \bibinfo{journal}{\bibinfo{title}{Fast unfolding of communities in
  large networks}}.
\newblock {\emph{\JournalTitle{Journal of Statistical Mechanics: Theory and
  Experiment}}} \textbf{\bibinfo{volume}{2008}}, \bibinfo{pages}{P10008}
  (\bibinfo{year}{2008}).

\bibitem{deville2014dynamic}
\bibinfo{author}{Deville, P.} \emph{et~al.}
\newblock \bibinfo{journal}{\bibinfo{title}{Dynamic population mapping using
  mobile phone data}}.
\newblock {\emph{\JournalTitle{Proceedings of the National Academy of
  Sciences}}} \textbf{\bibinfo{volume}{111}}, \bibinfo{pages}{15888--15893}
  (\bibinfo{year}{2014}).

\bibitem{barbosa2018human}
\bibinfo{author}{Barbosa, H.} \emph{et~al.}
\newblock \bibinfo{journal}{\bibinfo{title}{Human mobility: Models and
  applications}}.
\newblock {\emph{\JournalTitle{Physics Reports}}}
  \textbf{\bibinfo{volume}{734}}, \bibinfo{pages}{1--74}
  (\bibinfo{year}{2018}).

\bibitem{moro2021mobility}
\bibinfo{author}{Moro, E.}, \bibinfo{author}{Calacci, D.},
  \bibinfo{author}{Dong, X.} \& \bibinfo{author}{Pentland, A.}
\newblock \bibinfo{journal}{\bibinfo{title}{Mobility patterns are associated
  with experienced income segregation in large us cities}}.
\newblock {\emph{\JournalTitle{Nature Communications}}}
  \textbf{\bibinfo{volume}{12}}, \bibinfo{pages}{1--10} (\bibinfo{year}{2021}).

\bibitem{gao2022quantifying}
\bibinfo{author}{Gao, T.} \emph{et~al.}
\newblock \bibinfo{journal}{\bibinfo{title}{Quantifying relation between
  mobility patterns and socioeconomic status of dockless sharing-bike users}}.
\newblock {\emph{\JournalTitle{arXiv preprint arXiv:2204.03894}}}
  (\bibinfo{year}{2022}).

\bibitem{jiang2016timegeo}
\bibinfo{author}{Jiang, S.} \emph{et~al.}
\newblock \bibinfo{journal}{\bibinfo{title}{The timegeo modeling framework for
  urban mobility without travel surveys}}.
\newblock {\emph{\JournalTitle{Proceedings of the National Academy of
  Sciences}}} \textbf{\bibinfo{volume}{113}}, \bibinfo{pages}{E5370--E5378}
  (\bibinfo{year}{2016}).

\bibitem{yan2014universal}
\bibinfo{author}{Yan, X.-Y.}, \bibinfo{author}{Zhao, C.}, \bibinfo{author}{Fan,
  Y.}, \bibinfo{author}{Di, Z.} \& \bibinfo{author}{Wang, W.-X.}
\newblock \bibinfo{journal}{\bibinfo{title}{Universal predictability of
  mobility patterns in cities}}.
\newblock {\emph{\JournalTitle{Journal of the Royal Society Interface}}}
  \textbf{\bibinfo{volume}{11}}, \bibinfo{pages}{20140834}
  (\bibinfo{year}{2014}).

\bibitem{simini2012universal}
\bibinfo{author}{Simini, F.}, \bibinfo{author}{Gonz{\'a}lez, M.~C.},
  \bibinfo{author}{Maritan, A.} \& \bibinfo{author}{Barab{\'a}si, A.-L.}
\newblock \bibinfo{journal}{\bibinfo{title}{A universal model for mobility and
  migration patterns}}.
\newblock {\emph{\JournalTitle{Nature}}} \textbf{\bibinfo{volume}{484}},
  \bibinfo{pages}{96--100} (\bibinfo{year}{2012}).

\bibitem{wang2015beijing}
\bibinfo{author}{Wang, H.} \emph{et~al.}
\newblock \bibinfo{journal}{\bibinfo{title}{Beijing passenger car travel
  survey: implications for alternative fuel vehicle deployment}}.
\newblock {\emph{\JournalTitle{Mitigation and Adaptation Strategies for Global
  Change}}} \textbf{\bibinfo{volume}{20}}, \bibinfo{pages}{817--835}
  (\bibinfo{year}{2015}).

\bibitem{park2018generalized}
\bibinfo{author}{Park, H.~J.}, \bibinfo{author}{Jo, W.~S.},
  \bibinfo{author}{Lee, S.~H.} \& \bibinfo{author}{Kim, B.~J.}
\newblock \bibinfo{journal}{\bibinfo{title}{Generalized gravity model for human
  migration}}.
\newblock {\emph{\JournalTitle{New Journal of Physics}}}
  \textbf{\bibinfo{volume}{20}}, \bibinfo{pages}{093018}
  (\bibinfo{year}{2018}).

\bibitem{yan2017universal}
\bibinfo{author}{Yan, X.-Y.}, \bibinfo{author}{Wang, W.-X.},
  \bibinfo{author}{Gao, Z.-Y.} \& \bibinfo{author}{Lai, Y.-C.}
\newblock \bibinfo{journal}{\bibinfo{title}{Universal model of individual and
  population mobility on diverse spatial scales}}.
\newblock {\emph{\JournalTitle{Nature Communications}}}
  \textbf{\bibinfo{volume}{8}}, \bibinfo{pages}{1--9} (\bibinfo{year}{2017}).

\bibitem{feng2018evolving}
\bibinfo{author}{Feng, M.}, \bibinfo{author}{Deng, L.} \&
  \bibinfo{author}{Kurths, J.}
\newblock \bibinfo{journal}{\bibinfo{title}{Evolving networks based on birth
  and death process regarding the scale stationarity}}.
\newblock {\emph{\JournalTitle{Chaos: An Interdisciplinary Journal of Nonlinear
  Science}}} \textbf{\bibinfo{volume}{28}}, \bibinfo{pages}{083118}
  (\bibinfo{year}{2018}).

\bibitem{li2021assessing}
\bibinfo{author}{Li, R.} \emph{et~al.}
\newblock \bibinfo{journal}{\bibinfo{title}{Assessing the attraction of cities
  on venture capital from a scaling law perspective}}.
\newblock {\emph{\JournalTitle{IEEE Access}}} \textbf{\bibinfo{volume}{9}},
  \bibinfo{pages}{48052--48063} (\bibinfo{year}{2021}).

\bibitem{bak2013nature}
\bibinfo{author}{Bak, P.}
\newblock \emph{\bibinfo{title}{How nature works: the Science of self-organized
  criticality}} (\bibinfo{publisher}{Springer Science \& Business Media},
  \bibinfo{year}{2013}).

\bibitem{christensen2005complexity}
\bibinfo{author}{Christensen, K.} \& \bibinfo{author}{Moloney, N.~R.}
\newblock \emph{\bibinfo{title}{Complexity and criticality}},
  vol.~\bibinfo{volume}{1} (\bibinfo{publisher}{World Scientific Publishing
  Company}, \bibinfo{year}{2005}).

\bibitem{ke2015defining}
\bibinfo{author}{Ke, Q.}, \bibinfo{author}{Ferrara, E.},
  \bibinfo{author}{Radicchi, F.} \& \bibinfo{author}{Flammini, A.}
\newblock \bibinfo{journal}{\bibinfo{title}{Defining and identifying sleeping
  beauties in science}}.
\newblock {\emph{\JournalTitle{Proceedings of the National Academy of
  Sciences}}} \textbf{\bibinfo{volume}{112}}, \bibinfo{pages}{7426--7431}
  (\bibinfo{year}{2015}).

\bibitem{doumbia2018emissions}
\bibinfo{author}{Doumbia, M.} \emph{et~al.}
\newblock \bibinfo{journal}{\bibinfo{title}{Emissions from the road traffic of
  west african cities: Assessment of vehicle fleet and fuel consumption}}.
\newblock {\emph{\JournalTitle{Energies}}} \textbf{\bibinfo{volume}{11}},
  \bibinfo{pages}{2300} (\bibinfo{year}{2018}).

\bibitem{jiang2013review}
\bibinfo{author}{Jiang, S.} \emph{et~al.}
\newblock \bibinfo{title}{A review of urban computing for mobile phone traces:
  current methods, challenges and opportunities}.
\newblock In \emph{\bibinfo{booktitle}{Proceedings of the 2nd ACM SIGKDD
  international workshop on Urban Computing}}, \bibinfo{pages}{1--9}
  (\bibinfo{year}{2013}).

\bibitem{zheng2011learning}
\bibinfo{author}{Zheng, Y.} \& \bibinfo{author}{Xie, X.}
\newblock \bibinfo{journal}{\bibinfo{title}{Learning travel recommendations
  from user-generated gps traces}}.
\newblock {\emph{\JournalTitle{ACM Transactions on Intelligent Systems and
  Technology (TIST)}}} \textbf{\bibinfo{volume}{2}}, \bibinfo{pages}{1--29}
  (\bibinfo{year}{2011}).

\bibitem{guttman1984r}
\bibinfo{author}{Guttman, A.}
\newblock \bibinfo{title}{R-trees: A dynamic index structure for spatial
  searching}.
\newblock In \emph{\bibinfo{booktitle}{Proceedings of the 1984 ACM SIGMOD
  international conference on Management of data}}, \bibinfo{pages}{47--57}
  (\bibinfo{year}{1984}).

\bibitem{xu2019unraveling}
\bibinfo{author}{Xu, Y.} \emph{et~al.}
\newblock \bibinfo{journal}{\bibinfo{title}{Unraveling environmental justice in
  ambient pm2. 5 exposure in beijing: A big data approach}}.
\newblock {\emph{\JournalTitle{Computers, Environment and Urban Systems}}}
  \textbf{\bibinfo{volume}{75}}, \bibinfo{pages}{12--21}
  (\bibinfo{year}{2019}).

\bibitem{jingfang2015folding}
\bibinfo{author}{Hao, J.}
\newblock \bibinfo{journal}{\bibinfo{title}{Folding beijing}}.
\newblock {\emph{\JournalTitle{Translated by Ken Liu, Uncanny Magazine}}}
  \textbf{\bibinfo{volume}{2}} (\bibinfo{year}{2015}).

\end{thebibliography}

\end{document}